\documentclass[aps,prb,twocolumn,longbibliography]{revtex4-2}
\usepackage{graphicx,amsmath,amssymb,bm}

\pdfoutput=1
\usepackage[caption=false]{subfig}
\usepackage{array}
\usepackage{slashed,bbold}
\usepackage{physics}
\usepackage{dsfont}

\usepackage{amsmath,amssymb,bm} 
\usepackage{graphicx}
\usepackage{soul}
\usepackage{xcolor}
\usepackage{colortbl}
\usepackage{color} 
\definecolor{darkblue}{rgb}{0.,0.,0.4}
\definecolor{darkred}{rgb}{0.5,0.,0.}
\definecolor{BlueViolet}{RGB}{138,43,226}
\definecolor{SkyBlue}{RGB}{30,144,255}
\definecolor{DarkGreen}{RGB}{0,100,0}
\usepackage[pdftex,colorlinks=true,linkcolor=darkblue,citecolor=blue,urlcolor=darkred]{hyperref}

\begin{document}
	

\title{ Role of isotropic and anisotropic Dzyaloshinskii-Moriya interaction on skyrmions, merons and antiskyrmions in the $C_{nv}$ symmetric system} 
\author{Sandip Bera }
\affiliation{Department of Physics, University of Toronto, 60 St. George Street, Toronto, Ontario, Canada M5S 1A7	}	 
	
\begin{abstract}
	
The lattice Hamiltonian with the presence of a chiral magnetic isotropic Dzyaloshinskii-Moriya interaction (DMI) in a square and hexagonal lattice is numerically solved to give the full phase diagram consisting of skyrmions and merons in different parameter planes.  The phase diagram provides the actual regions  of analytically unresolved asymmetric skyrmions and merons, and it is found that these regions are substantially larger than those of symmetric skyrmions and merons.
With magnetic field, a change from meron or spin spiral to skyrmion is seen. The complete phase diagram for the $C_{nv}$ symmetric system with anisotropic DMI is drawn and it is shown that this DMI helps to change the spin spiral propagation direction.	 Finally, the well-defined region of a thermodynamically stable antiskyrmion phase in the $C_{nv}$ symmetric system  is shown.

\end{abstract}

\maketitle

The Dzyaloshinskii-Moriya interaction \cite{Dzyaloshinskii1957,Dzyaloshinskii1965,Moriya1960}, an asymmetric exchange interaction, plays an important role in stabilizing the noncollinear magnetic structure in broken inversion symmetric systems. The spin spiral (SS), whose length scale \cite{Bogdanov1989,Hubert1994,Leonov2016,sandip2019} is mostly created by DMI and symmetric exchange interactions, is a common noncollinear ground state of a chiral  magnet.  This ground state transforms into a number of interesting topologically protected small-scale spin textures, including skyrmions  \cite{Meyer2019,Munzer2010,Yu2016,Romming2013,Yu2010,Herve2018,Yu2018} and merons \cite{Yu2018564,Gao2019}, with the aid of magnetic fields, magnetic anisotropy, and higher order frustrated exchange interactions. These spin textures are classified into two types based on crystal symmetry: Neel type for $C_{nv}$ symmetric and Block type for $D_n$ symmetric systems \cite{Bogdanov1989,Hubert1994}. Whereas, the $D_{2d}$ symmetric system \cite{Leonov2016} supports antiskyrmions \cite{Nayak2017} and antimerons \cite{Hayami2021}. Due to their tiny size \cite{Nagaosa2013,Soumyanarayanan2017}, topological stability\cite{Bogdanov1989,Hubert1994}, and stability over a broad range of parameter spaces \cite{sandip2019,sandip2020,Leonov2016,Banerjee2014,Nandy2016} including magnetic field, anisotropy, and DMI, all of these topological non-trivial phases have had a positive impact on the field of spintronics\cite{Parkin2008,Iwasaki2013,Fert2013,Yu2017,Sampaio2013,Jonietz2010}.

We have a semi-analytic theory \cite{Hubert1994,sandip2019,Leonov2016} for singly symmetric skyrmions in a ferromagnetic background.  However, we still have no analytical solution for experimentally  observed asymmetric skyrmions \cite{Huang2012,Li2013,Porter2014,Yokouchi2014}  in the presence of easy plane anisotropy. Since the topological number remains the same  \cite{Gobel2021,sandip2019}, asymmetric skyrmions have the same topological properties as symmetric skyrmions, such as the Hall effect \cite{sandip2020l}. Asymmetric merons such as asymmetric skyrmions have been observed experimentally  \cite{Yu2018564,Shigeto2002,Chmiel2018,Phatak2012} in regions of easy plane anisotropy. Even merons always appear in lattice form \cite{Yu2018564,Gao2019,sandip2019}. Continuum model-based analytical solutions, however, are more suitable for single symmetric skyrmions and merons. As a result, we are unable to construct precise phase diagrams for the lattice skyrmions, asymmetric skyrmions,  symmetric merons and  asymmetric merons using the continuum model. In this article, I describe the full phase diagram by minimizing the lattice Hamiltonian in presence of isotropic DMI  using Monte Carlo (MC) simulations with periodic boundary conditions on both square and hexagonal lattices. These phase diagrams depict  a transition from SS  $\rightarrow$ skyrmions (lattice skyrmion, isolated skyrmion)  $\rightarrow$ polar ferromagnet in the easy-axis anisotropy limit and from merons  $\rightarrow$ skyrmions in the easy-plane anisotropy limit, all of which have already been observed experimentally \cite{Yu2010,Herve2018,Iwasaki2013,Yu2018}.
These phase  diagrams also demonstrate that asymmetric skyrmions and asymmetric merons have significantly wider regions in the parameter space than symmetric skyrmions and symmetric merons.

The orthogonal coefficients of an anisotropic DMI are different. Heide et al. \cite{Heide2008}  predicted such DMI in  2Fe/W(100) films  using first-principles calculations. It has been observed experimentally in  an ultrathin epitaxial Au/Co/W(110) films \cite{Camosi2017}, CoFeB(211)/Pt(110) films \cite{Liu2021}.    Shibata et al.  \cite{Shibata2015} showed that the anisotropy of bulk DMI causes mechanical strain, which produces an anisotropic skyrmion. We already know that anisotropic DMI is direction dependant, and this strain aids in changing the spin configuration. This DMI can also alter the spin orientation of the skyrmions in a certain manner, leading to the eventual emergence of an antiskyrmions  in $C_{nv}$ symmetric system. Hoffmann et al. \cite{Hoffmann2017} have  first observed antiskyrmions in (110)-oriented films with $C_{2v}$  symmetric systems. Although, we generally know skyrmions appears in $C_{2v}$ symmetric systems. We have a theoretical prediction \cite{sandip2019} of antiskyrmion stabilization in $C_{nv}$ symmetric systems. However, neither this skyrmion nor the antiskyrmion is symmetric about their center. Consequently, there are no analytical solutions and phase diagrams for these topological phases. Here, I first demonstrate through MC simulations how the anisotropic DMI promotes the skyrmion to antiskyrmion transition in the $C_{nv}$ symmetric system before illuminating the full phase diagram for this DMI.

\section{ Hamiltonian and Methods} 
The ferromagnetic Heisenberg spin Hamiltonian in the presence of an easy axis magnetic field can be written as

\begin{equation}\label{j1_lattice_model}
{\cal H} = -J\sum_{<ij>}{\bf m}_i \cdot {\bf m}_j+H_{DMI}+A\sum_{i}(m^z_{i})^2 -H\sum_{i} {m}^z_{i},
\end{equation}
where ${\bf m}_{i}$ is the unit vector of the magnetization at $i^{th}$ sites. The first term is the  isotropic  exchange interaction with nearest-neighbor coupling constant, $J$. Second term,  $H_{DMI}$ is an  antisymmetric spin orbit interaction, known as Dzyaloshinskii-Moriya interaction \cite{Dzyaloshinskii1957}. The direction  of the DMI vectoer depends on the crystal symmetries.  $H_{DMI}$ is proportional to $D\sum_{<ij>} (\hat{z}\times \hat{r}_{ij})\cdot({\bf m}_i \times {\bf m}_j)$   for $C_{nv}$, and $D\sum_{<ij>} \hat{r}_{ij}\cdot({\bf m}_i \times {\bf m}_j)$   for $D_{n}$ symmetries  systems respectively. Here, $\hat{r}_{ij}$ is unit vector from $i$ to $j$ sites. The amplitude of DMI vector ($D$) is proportional to interaction strength between spin and orbit. Here, $H_{DMI}$ is isotropic.  $A$  is the coefficient of the  magnetic anisotropy. It is positive for easy-plane anisotropy and negative for easy-axis anisotropy. $H$ is the effective magnetic field perpendicular to the plane of the systems.  The last two terms  in Eq. \eqref{j1_lattice_model} determine the direction of the ferromagnet \cite{Banerjee2014,sandip2019}. When  $A<0$, all spins align parallel to $H$ and form a polar ferromagnet phase. For $A>0$ , the direction of magnetization depends on the anisotropy and the magnetic field strength. Easy-plane anisotropy forces all spins to lie in the plane, and the magnetic field forces the spins to lie perpendicular to the plane. Consequently all spins make an angle with the direction of $H$. This configuration is known as tilted ferromagnet.


The Hamiltonian given in Eq. \eqref{j1_lattice_model} is minimised on both the square lattice and the hexagonal lattice using a MC simulation based on the conventional Metropolis algorithm \cite{Metropolis1953,Hayami2018} with periodic boundary conditions.  Hamiltonian has many local minima or metastable state for a given parameters. To avoid these metastable states and locate the actual thermodynamic ground state, conventional simulated annealing techniques with MC are employed. The MC simulation begins with a random initial spin configuration at a high temperature. The temperature is then steadily lowered using the formula $T_{n+1}=\alpha T_{n}$, with an MC simulation being run at each stage.  The initial spin configuration in the MC simulation is derived from the outcome of the preceding step after the first annealing loop has been completed.  Repeat the same procedure until my final temperature reaches a very low value ( close to zero). The simulation parameters used in my numerical calculations are: total  number of steps in annealing method $n=10^3$, initial temperature $T_{0}=11 J$,  final  temperature $\approx 5\times 10^{-4}J $, $\alpha=0.99$, and MC steps $4\times 10^{6}$. $J$ is utilized to measure all energy-related parameters such as   $D$, $A$ and $H$, while $a$, the lattice constant, is used to scale length. $D=\sqrt{6}J$ is chosen to construct the phase diagram in the $H-A$ plane. This value of DMI supports a SS with periodicity close to $8a$.  We know that the periodicity of SS in $MnGe$ is  about $8.3a$ \cite{Kanazawa2011}. So, the chosen value of DMI belongs to real materials. All these MCs are performed  $32\times 32$  mesh, but the mesh size is increased to $64\times 64$ for some important results (mostly at phase boundaries). My simulation always treats  $J$ as 1.

\subsection{Numerical Results and Phase Diagrams in $C_{nv}$ Symmetric System}

The Hamiltonian \eqref{j1_lattice_model} has only three parameters, namely $D$, $H$ and $A$ because $J$ is used to scale all energy parameters. We only have the $D-H$, $D-A$, and $H-A$ planes for the phase diagram.  The  article is organized as follows: i  begin by showing the whole phase diagram for the $C_{nv}$ symmetric system with isotropic DMI on a square and a hexagonal lattice of different parameter planes.  The numerical outcomes for the $D_{2d}$ symmetric system in $D-H$ plane are later explained. Finally, in the presence of anisotropic DMI, numerical results and an entire phase diagram are displayed for the $C_{nv}$ symmetric system.

The final output of the MC gives a spin configuration of the system for a given parameter. To identify the phase that this spin configuration supports, we need to use some distinct properties: local magnetization $m^i_z$, average magnetization $M=1/N\sum_{i}m^i_z$, local chirality $\rho_i = (1/4\pi) \sum_{\delta=\pm 1} {\bf m}_i\cdot({\bf m}_{i+\delta\hat{x}}\times {\bf m}_{i+\delta\hat{y}})$, and total chirality $\mathcal{R} = (1/N)\sum_{i} \rho_i $, local spin asymmetric parameter $\delta\theta_{i} = |\theta_{i+\hat{y}} -\theta_{i-\hat{y}}|-|\theta_{i+\hat{x}}-\theta_{i-\hat{x}}| $, where N is the number of lattice sites. $\delta\theta_{i}$ gives information about symmetry with respect to its center. Chirality helps distinguish between topologically trivial and nontrivial phases. A detailed description of these above parameters for different phases is given in Ref. \cite{sandip2020}. These characteristic variables are used to draw phase boundaries in all phase diagrams.

\begin{figure}[h]
  	\begin{center}	
  		\subfloat[\label{phase_square}]{\includegraphics[scale=0.24]{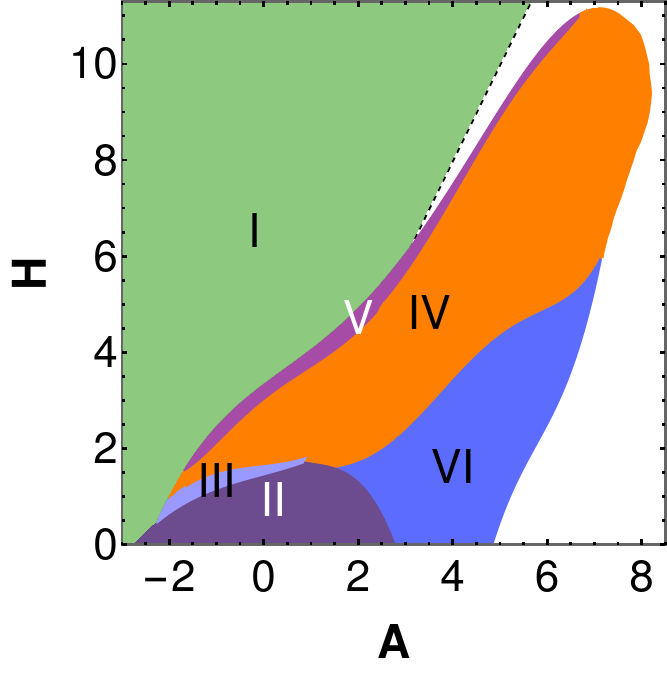}}~
  		\subfloat[\label{phase_meron}]{	\includegraphics[scale=0.24]{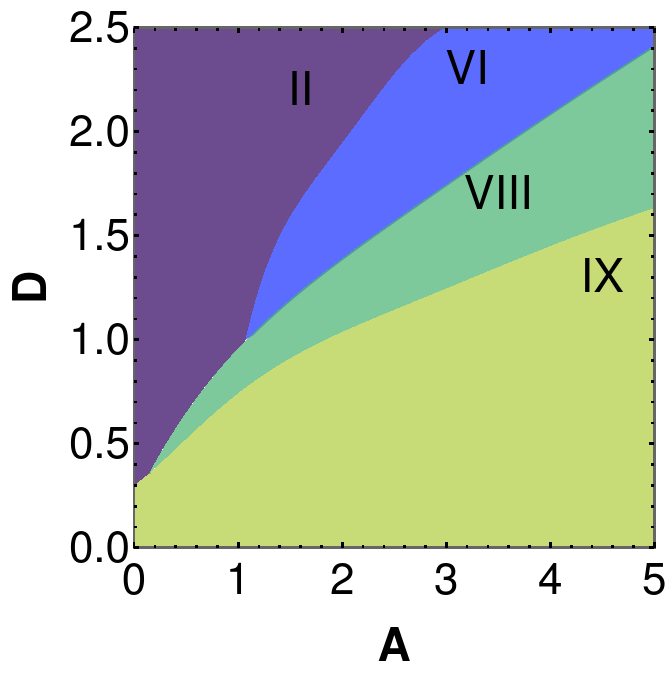}}~
  		\subfloat[\label{phase_hexagonal}]{\includegraphics[scale=0.23]{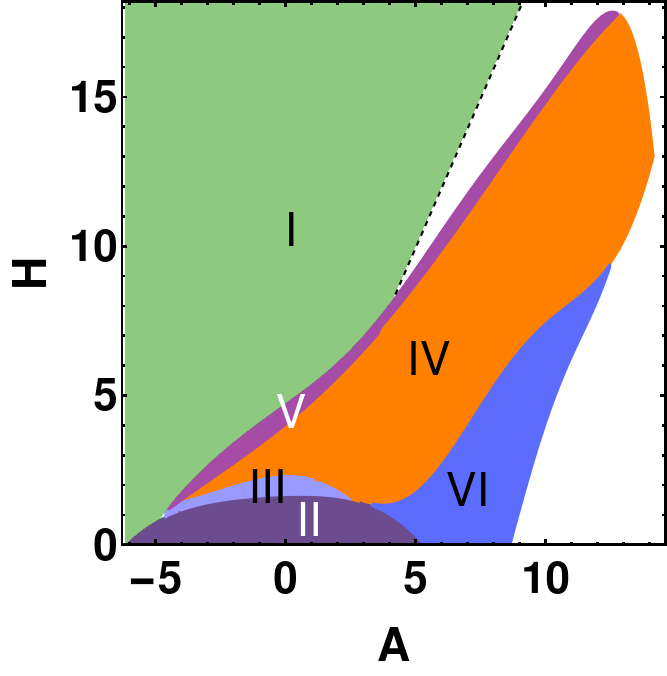}}~
  		\subfloat[\label{antiskyrmionphase}]{\includegraphics[scale=0.24]{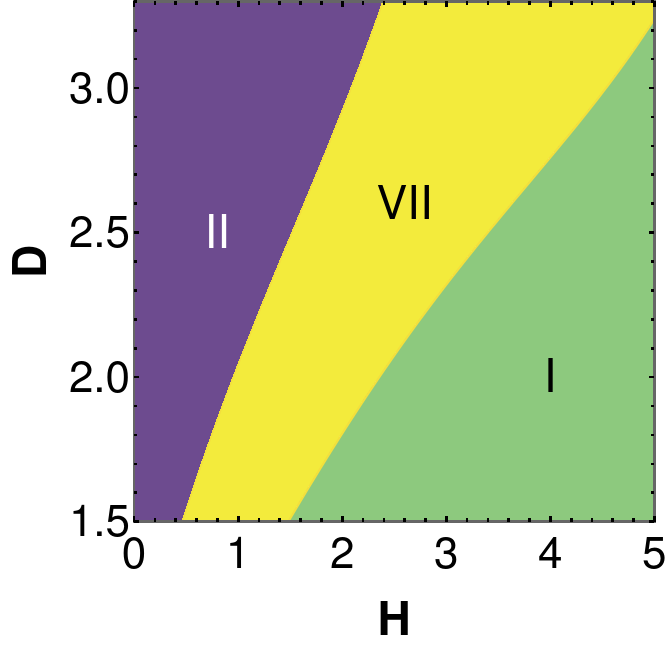}}	
  		\caption{  Phase diagrams: \eqref{phase_square} $H-A$ plane at $D=\sqrt{6},~ J=1$   and \eqref{phase_meron} $D-A$ plane at $H=0,~ J=1$. The important phases in the phase diagrams are: Ferromagnet  with magnetization along $\hat{z}$ direction (I),  spin spiral (II), mixed phase of broken spin spiral and skyrmion (III), skyrmion lattice (IV), isolated skyrmion (V), meron (VI), Planar FM (IX) and one unknown phase (VIII).   It is challenging to discern between the phase-VIII and the tilted FM phase in my numerical simulations. So, in the phase diagrams (Figures \ref{phase_square} and \ref{phase_hexagonal}, I have highlighted them in white without suitable borders.  The complete phase diagram for a hexagonal lattice at $D=\sqrt{6},~ J=1$ is displayed in panel \eqref{phase_hexagonal} in the $H-A$ plane.   Panel (d) shows the phase diagram on a square lattice in the $D_{2d}$ system at $A=0$. Here, the antiskyrmion (VII) phase region grew with $D$ and $H$. The spin configurations of all these phases are shown in figure \eqref{lambda_anisotropy_all}.} 
  		\label{phase_square_all}
  	\end{center}
  \end{figure}

\begin{figure}[h]
  	\begin{center}
  		\includegraphics[trim={0.5cm 3.5cm 1.0cm 2.6cm},clip,scale=0.54]{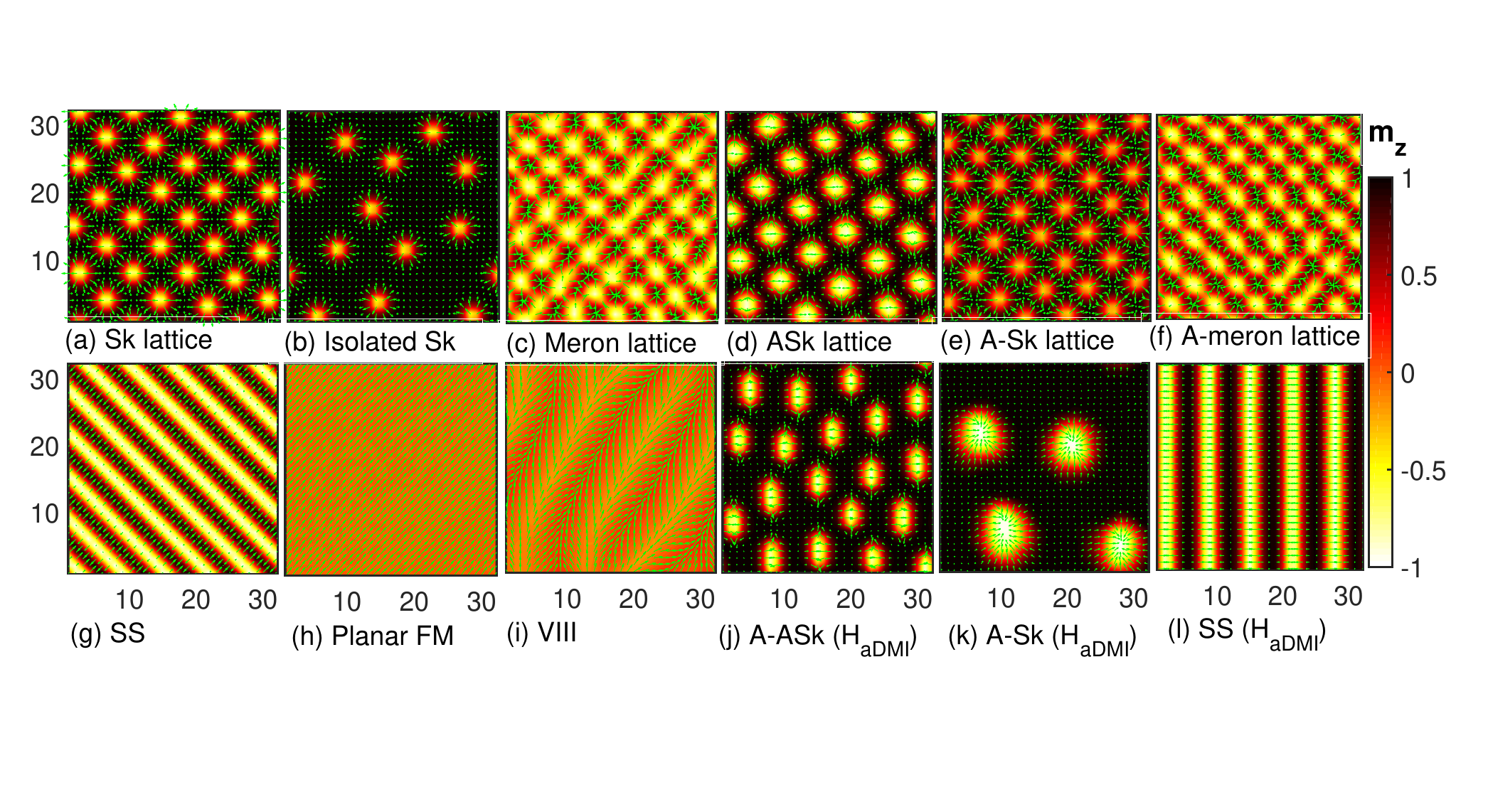}	
  		\caption{Magnetic structures obtained  from MC  simulations for different phases are shown in panels: (a) skyrmion (Sk) lattice, (b) Isolated skyrmion, (c) meron lattice, (d) antiskyrmion (ASk) lattice, (e) asymmetric skyrmion (A-Sk) lattice, (f) Asymmetric meron (A-meron) lattice, (g) spin spiral (SS), (h) Ploar ferromagnet (FM), (i) unknown phase (VIII), (j) asymmetric antiskyrmion (A-ASk) in presence of anisotropic DMI ($H_{aDMI}$), (k) asymmetric skyrmion (A-Sk) in presence of $H_{aDMI}$, (l) Spin spiral in presence of $H_{aDMI}$. The out-of-plane component of magnetization is color coded, while the in-plane component is shown by arrows.} 
  		\label{lambda_anisotropy_all}
  	\end{center}
  \end{figure}

The complete phase diagram for a square lattice in the $H-A$ plane at $D=\sqrt{6}$ is shown in Fig. \eqref{phase_square}.  The large magnetic field transforms the system into the ferromagnetic (FM) phase (Phase-1). In opposite limit, a weak magnetic field causes the spins to form a spin spiral (SS) phase (phase-II). Here, SS propagates along the $x+y$ direction because the sign of the DMI vector is taken positive. These two phases, SS and FM, are topologically trivial and emerge at extreme magnetic field values.  Now increasing the magnetic field in the SS phase in the easy-plane anisotropy region. At this stage, all spins begin to experience an additional force in the presence of a magnetic field. The magnetic field seeks to align all of the spins in the field's direction. DMI, however, aims to provide a finite rotation between the spins.  These interactions work together to stabilize a skyrmion. At low field ranges, the skyrmion and spin spiral appear together in what is known as a mixed phase (Phase-III). As the magnetic field increases, the number of spin spirals decreases. Because of this, we've reached a point where there are only skyrmions in the system. In this case, the separation between any two skyrmions remains the same and forms a lattice known as  skyrmion lattice (phase-IV).  The magnetic field increases the spacing between the skyrmions. As a result, fewer skyrmions exist in a given lattice shape and they are also smaller in size. These small-sized skyrmions are randomly distributed in what is called the isolated skyrmion phase (phase-V). The decrease in size of the skyrmions with magnetic field has already been predicted theoretically \cite{sandip2019} and observed experimentally \cite{Romming2015}. The area of phase-V is much smaller than the phase-IV. Finally for a slight increase in the magnetic field we reach the FM phase. The spin configurations of skyrmion lattice phase and  isolated skyrmion phase  are shown in figures (\ref{lambda_anisotropy_all}a) and (\ref{lambda_anisotropy_all}b), respectively.  We can see that there is a transition from SS to skyrmion to FM with magnetic field which has already been observed experimentally \cite{Yu2010,Herve2018,Iwasaki2013,Yu2018}.

\onecolumngrid
 
\vspace{-0.3cm}  \vspace{-0.6cm}
  %
  %
  %
  %
  %
  \vspace{-0.5cm}
  
\twocolumngrid 
  
The phase diagram at $A>0$ is interesting because there are no analytical solutions for the asymmetric skyrmions and asymmetric merons in this region.  Figure \eqref{phase_square} shows the corresponding phase diagram. We observed both skyrmions and SS phases here, similar to negative anisotropy.  Here $\delta\theta_{i} \neq 0$ for all phases, indicating that these skyrmions are not symmetric about their center called asymmetric skyrmions. This result is consistent with analytical predictions \cite{sandip2019}.  The phase space region of these skyrmions is larger in this case than it is in the $A<0$ case, which is a second crucial fact. Given the wide spectrum of asymmetric skyrmion, this is crucial in application. Spin configuration of a asymmetric skyrmion  is shown in figure (\ref{lambda_anisotropy_all}e).

Apart from the existing phases described in the above paragraph, there is another interesting phase which is also  topologically non-trivial known as the meron phase (phase-VI). This phase lies below the skyrmion-lattice phase. Like skyrmions, merons are not symmetrical about their center in regions known as asymmetric merons. Spin configuration  of asymmetric merons  is shown in figure (\ref{lambda_anisotropy_all}f). However, at $H=0$ the merons are formed in the background of planar ferromagnet and are symmetric about their center. Figure \eqref{phase_square} shows that $2.4-4.5$ is the range of anisotropy for symmetric merons at $H=0$. Both types of merons appear in lattice shapes that are consistent with theoretical predictions \cite{Yu2018564,Gao2019,sandip2019}. This result is new and interesting not only because it is challenging to estimate the potential meron region using continuum models, but also because these results are crucial for practical applications. Phases in the white area in Fig. \eqref{phase_square} are difficult to identify because this region may contain some conical phases and tilted ferromagnet.

Fig. \eqref{phase_meron} shows the full phase diagram in the $D-A$ plane for a better understanding of a symmetric meron.  The phase region of symmetric merons increases with  both DMI and magnetic anisotropy. This phase diagram also suggests that we need a DMI value close to the exchange interaction strength to stabilize symmetric merons in the absence of a magnetic field. Which indicates that the system should have strong spin-orbit coupling to support symmetric merons. The planar ferromagnetic phase (phase IX) is observed with large anisotropy. An unknown phase (phase VIII) is noticed between merons and planar ferromagnetic phases. The spin configuration of these phases are shown in Fig. (\ref{lambda_anisotropy_all}).

The full phase diagram for the hexagonal lattice is shown in figure \eqref{phase_hexagonal}. The order and appearance of the phases  in the phase diagram are almost the same as in the square lattice phase diagram. However, in the hexagonal case the boundaries of each phase are slightly larger. In other words, each axis of our phase diagram has an additional scaling factor. Here, I consider $2a$ to be the lattice spacing for the hexagonal lattice (where $a$ is the lattice spacing for the square lattice). This additional scaling factor of the phase diagram is due to this lattice spacing.

\subsection{Antiskyrmion in $D_{2d}$ Symmetric System}
The same numerical method is used to minimize the lattice Hamiltonian (\ref{j1_lattice_model}) in $D_{2d}$ symmetric systems for square lattice. Here the phase diagram is similar to the $C_{nv}$ symmetric system, but the $D_{2d}$ symmetric system has antiskyrmions instead of skyrmions. Since the results are the same, we have calculated the phase diagram in a different plane ($D-H$) for the $D_{2 d}$ systems. Fig. \eqref{antiskyrmionphase} shows that the antiskyrmions phase (phase-VII) regions increase with increasing DMI and magnetic field.  Spin configuration of an antiskyrmion is shown in Fig. (\ref{lambda_anisotropy_all}d).

\begin{figure}
	\begin{center}
		\includegraphics[scale=0.36]{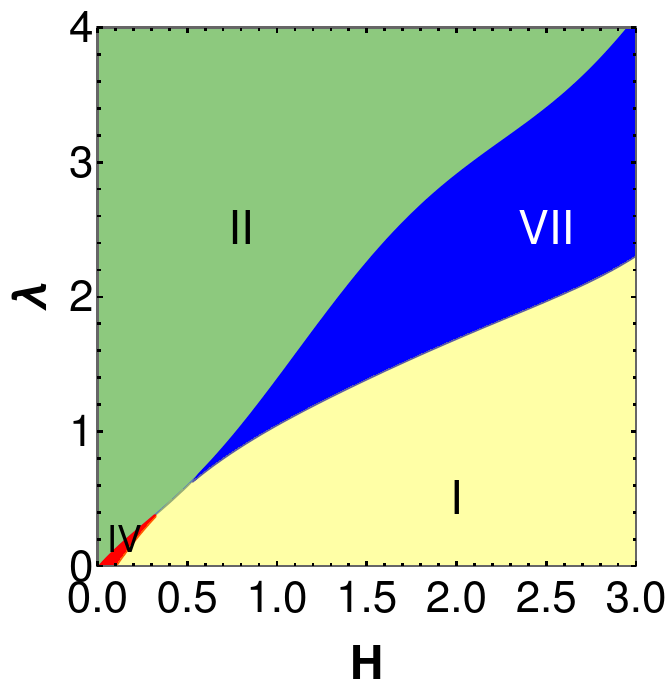}		
		\caption{ Phase  diagram  with an anisotropic  DMI in $C_{nv}$ symmetric system at $A=0$. Here, the spin configurations (see Fig. \eqref{lambda_anisotropy_all}) of spin spiral (II), skyrmion (IV) and antiskyrmion (VII) are slightly different with the same phases observed in the case of isotropic DMI. } 
		\label{phase_sk_ask}
	\end{center}
\end{figure}
\section{Antiskyrmion in $C_{nv}$ Symmetric System with Anisotropic DMI }

The following  form of anisotropic DMI ($H_{aDMI}$)   is considered in the lattice Hamiltonian:

\begin{align}\label{ani_equation}
H_{aDMI}= (1+\lambda)D_{x}+ (1-\lambda)D_{y},
\end{align}
where $\lambda$ is a positive number and represents the strength of the anisotropic DMI. The DMI becomes isotropic when $\lambda=0$.  Anisotropic DMI refers to the fact that the DMI's amplitude in the $x$ and $y$ directions differ. Here, $D_x$ and $D_y$ are, in the case of an isotropic DMI vector, the $x$ and $y$ components, respectively. The introductory section has already described how this term is expressed for several symmetric systems.  In Eq. \eqref{ani_equation}, the same amount of anisotropy added to the $x$ component of $H_{aDMI}$ is subtracted from the $y$ component.

Here, my attention is solely on  $C_{nv}$ symmetric system. It is obvious that all skyrmions and merons will be asymmetric about their centres because our DMI is asymmetric in nature. The entire phase diagram at $A=0$ is displayed in Fig. \eqref{phase_sk_ask} in the $\lambda-H$ plane. The SS phase (phase II) manifests at large anisotropic DMI or large  $\lambda$, as expected. The spin configuration of SS phase  in this instance differs from the  spin configuration  of SS phase in isotropic DMI (Fig. (\ref{lambda_anisotropy_all}l) ).  For $\lambda=0.72$ and $H=1.0$ SS propagates along $x-$direction whereas SS in isotropic DMI always propagates along $x+y$-direction (Because the isotropic DMI vector is assumed to be a positive quantity).  This result demonstrates that we can continually modify the SS phase propagation direction by utilising $\lambda$. In the other limit, a ferromagnetic phase (I-phase) is observed at very low $\lambda$ and large $H$. Skyrmions (phase-IV) are observed in a small region with low magnetic field and low $\lambda$. These skyrmions are asymmetrical and extended in the $y$ direction. For any $\lambda$, the $y$ component of $H_{aDMI}$  is smaller than the $x$ component.  It is known that the radii \cite{sandip2019,Leonov2016} of  skyrmions and merons depend inversely on the DMI. Thus, the extension of the skyrmions border along the $y$ direction seen in our simulations is also supported by the theoretical realization. Fig. (\ref{lambda_anisotropy_all}k)  depicts the spin arrangement of these asymmetric skyrmions. Finally, for medium to high values of $H$ and $\lambda$, the phase with the greatest interest (phase-IV) is observed.  It is an antiskyrmion according to all investigations including topological number, spin configuration and chirality. This antiskyrmion is also not symmetric, known as asymmetric antiskyrmion, and is thermodynamically stable. This phase exists between the ferromagnetic phase and the SS phase. Its spin configuration   is shown in Fig.  (\ref{lambda_anisotropy_all}j). The phase space region of this asymmetric antiskyrmion expands as $H$ and $\lambda$ increase.  As a result, we see experimentally \cite{Hoffmann2017} confirmed thermodynamically stable antiskyrmions in $C_{nv}$   symmetric system with anisotropic DMI. Using the same form of anisotropic DMI, one can also stabilize the skyrmion in the $D_{2d}$ system. In this case $D_{x}$ and $D_{y}$ will depend on the DMI form of the $D_{2d}$ symmetric system.

In conclusion,  complete phase diagrams for skyrmions in both square and hexagonal lattices are shown. The phase diagram remains the same for these two different lattices.  The phase region of merons are seperately depicted in different plane. The numerical results for the total  phase diagram of the $D_{2d}$ symmetric system are the same as those of the $C_{nv}$ symmetric system. Skyrmions are not seen in this instance; antiskyrmions are. These results are consistent with theoretical predictions.
Symmetric and asymmetric skyrmions, merons and antiskyrmions have already been detected experimentally in many systems.  In comparison to symmetric skyrmions, asymmetric skyrmions have a larger parameter space region, which is significant for applications. 
Finally, it is shown that the $C_{nv}$ symmetric system has a thermodynamically stable antiskyrmion that is in line with the findings of an experiment\cite{Hoffmann2017}.  This anisotropic DMI  also helps to change the direction of spiral spiral propagation.
Higher order exchange interactions and frustrated exchange interactions are absent in our calculations. We know that exchange can stabilize the skyrmion even for very small DMI \cite{sandip2020}. Thus, frustrated high order exchange interaction and anisotropic DMI can give some additional features to the phase diagram.

\begin{acknowledgements}
I would like to thank Prof. Sudhansu Shekhar Mandal for initial discussions. I'm grateful to Abinash Kumar Shaw, Sahana Das and Sudipto Das for help in understanding Monte Carlo simulations. The initial work of the project was done using the computer facilities provided by the Department of Physics at IIT Kharagpur.
\end{acknowledgements}

\bibliography{ge_ref_copy}

\begin{thebibliography}{47}%
\makeatletter
\providecommand \@ifxundefined [1]{%
 \@ifx{#1\undefined}
}%
\providecommand \@ifnum [1]{%
 \ifnum #1\expandafter \@firstoftwo
 \else \expandafter \@secondoftwo
 \fi
}%
\providecommand \@ifx [1]{%
 \ifx #1\expandafter \@firstoftwo
 \else \expandafter \@secondoftwo
 \fi
}%
\providecommand \natexlab [1]{#1}%
\providecommand \enquote  [1]{``#1''}%
\providecommand \bibnamefont  [1]{#1}%
\providecommand \bibfnamefont [1]{#1}%
\providecommand \citenamefont [1]{#1}%
\providecommand \href@noop [0]{\@secondoftwo}%
\providecommand \href [0]{\begingroup \@sanitize@url \@href}%
\providecommand \@href[1]{\@@startlink{#1}\@@href}%
\providecommand \@@href[1]{\endgroup#1\@@endlink}%
\providecommand \@sanitize@url [0]{\catcode `\\12\catcode `\$12\catcode
  `\&12\catcode `\#12\catcode `\^12\catcode `\_12\catcode `\%12\relax}%
\providecommand \@@startlink[1]{}%
\providecommand \@@endlink[0]{}%
\providecommand \url  [0]{\begingroup\@sanitize@url \@url }%
\providecommand \@url [1]{\endgroup\@href {#1}{\urlprefix }}%
\providecommand \urlprefix  [0]{URL }%
\providecommand \Eprint [0]{\href }%
\providecommand \doibase [0]{https://doi.org/}%
\providecommand \selectlanguage [0]{\@gobble}%
\providecommand \bibinfo  [0]{\@secondoftwo}%
\providecommand \bibfield  [0]{\@secondoftwo}%
\providecommand \translation [1]{[#1]}%
\providecommand \BibitemOpen [0]{}%
\providecommand \bibitemStop [0]{}%
\providecommand \bibitemNoStop [0]{.\EOS\space}%
\providecommand \EOS [0]{\spacefactor3000\relax}%
\providecommand \BibitemShut  [1]{\csname bibitem#1\endcsname}%
\let\auto@bib@innerbib\@empty
\bibitem [{\citenamefont {Dzialoshinskii}(1957)}]{Dzyaloshinskii1957}%
  \BibitemOpen
  \bibfield  {author} {\bibinfo {author} {\bibfnamefont {I.}~\bibnamefont
  {Dzialoshinskii}},\ }\bibfield  {title} {\bibinfo {title} {Thermodynamic
  theory of "weak" ferromagnetism in antiferromagnetic substances},\
  }\href@noop {} {\bibfield  {journal} {\bibinfo  {journal} {Zh. Eksp. Theo.
  Fiz.}\ }\textbf {\bibinfo {volume} {32}},\ \bibinfo {pages} {1547} (\bibinfo
  {year} {1957})},\ \bibinfo {note} {[Sov. Phys. JETP 5, 1259
  (1957)]}\BibitemShut {NoStop}%
\bibitem [{\citenamefont {Dzialoshinskii}(1965)}]{Dzyaloshinskii1965}%
  \BibitemOpen
  \bibfield  {author} {\bibinfo {author} {\bibfnamefont {I.}~\bibnamefont
  {Dzialoshinskii}},\ }\bibfield  {title} {\bibinfo {title} {The theory of
  helicoidal structures in antiferromagnets. ii. metals},\ }\href@noop {}
  {\bibfield  {journal} {\bibinfo  {journal} {Sov. Phys. JETP}\ }\textbf
  {\bibinfo {volume} {20}},\ \bibinfo {pages} {223} (\bibinfo {year} {1965})},\
  \bibinfo {note} {[Sov. Phys. JETP 20, 223 (1965)]}\BibitemShut {NoStop}%
\bibitem [{\citenamefont {Moriya}(1960)}]{Moriya1960}%
  \BibitemOpen
  \bibfield  {author} {\bibinfo {author} {\bibfnamefont {T.}~\bibnamefont
  {Moriya}},\ }\bibfield  {title} {\bibinfo {title} {Anisotropic superexchange
  interaction and weak ferromagnetism},\ }\href
  {https://doi.org/10.1103/PhysRev.120.91} {\bibfield  {journal} {\bibinfo
  {journal} {Phys. Rev.}\ }\textbf {\bibinfo {volume} {120}},\ \bibinfo {pages}
  {91} (\bibinfo {year} {1960})}\BibitemShut {NoStop}%
\bibitem [{\citenamefont {Bogdanov}\ and\ \citenamefont
  {Yablonskii}(1989)}]{Bogdanov1989}%
  \BibitemOpen
  \bibfield  {author} {\bibinfo {author} {\bibfnamefont {A.~N.}\ \bibnamefont
  {Bogdanov}}\ and\ \bibinfo {author} {\bibfnamefont {D.~A.}\ \bibnamefont
  {Yablonskii}},\ }\href@noop {} {\bibfield  {journal} {\bibinfo  {journal}
  {Zh. Eksp. Teor. Fiz.}\ }\textbf {\bibinfo {volume} {95}},\ \bibinfo {pages}
  {178} (\bibinfo {year} {1989})}\BibitemShut {NoStop}%
\bibitem [{\citenamefont {Bogdanov}\ and\ \citenamefont
  {Hubert}(1994)}]{Hubert1994}%
  \BibitemOpen
  \bibfield  {author} {\bibinfo {author} {\bibfnamefont {A.}~\bibnamefont
  {Bogdanov}}\ and\ \bibinfo {author} {\bibfnamefont {A.}~\bibnamefont
  {Hubert}},\ }\bibfield  {title} {\bibinfo {title} {Thermodynamically stable
  magnetic vortex states in magnetic crystals},\ }\href
  {https://doi.org/https://doi.org/10.1016/0304-8853(94)90046-9} {\bibfield
  {journal} {\bibinfo  {journal} {Journal of Magnetism and Magnetic Materials}\
  }\textbf {\bibinfo {volume} {138}},\ \bibinfo {pages} {255} (\bibinfo {year}
  {1994})}\BibitemShut {NoStop}%
\bibitem [{\citenamefont {Leonov}\ \emph {et~al.}(2016)\citenamefont {Leonov},
  \citenamefont {Monchesky}, \citenamefont {Romming}, \citenamefont {Kubetzka},
  \citenamefont {Bogdanov},\ and\ \citenamefont {Wiesendanger}}]{Leonov2016}%
  \BibitemOpen
  \bibfield  {author} {\bibinfo {author} {\bibfnamefont {A.~O.}\ \bibnamefont
  {Leonov}}, \bibinfo {author} {\bibfnamefont {T.~L.}\ \bibnamefont
  {Monchesky}}, \bibinfo {author} {\bibfnamefont {N.}~\bibnamefont {Romming}},
  \bibinfo {author} {\bibfnamefont {A.}~\bibnamefont {Kubetzka}}, \bibinfo
  {author} {\bibfnamefont {A.~N.}\ \bibnamefont {Bogdanov}},\ and\ \bibinfo
  {author} {\bibfnamefont {R.}~\bibnamefont {Wiesendanger}},\ }\bibfield
  {title} {\bibinfo {title} {The properties of isolated chiral skyrmions in
  thin magnetic films},\ }\href
  {http://stacks.iop.org/1367-2630/18/i=6/a=065003} {\bibfield  {journal}
  {\bibinfo  {journal} {New Journal of Physics}\ }\textbf {\bibinfo {volume}
  {18}},\ \bibinfo {pages} {065003} (\bibinfo {year} {2016})}\BibitemShut
  {NoStop}%
\bibitem [{\citenamefont {Bera}\ and\ \citenamefont
  {Mandal}(2019)}]{sandip2019}%
  \BibitemOpen
  \bibfield  {author} {\bibinfo {author} {\bibfnamefont {S.}~\bibnamefont
  {Bera}}\ and\ \bibinfo {author} {\bibfnamefont {S.~S.}\ \bibnamefont
  {Mandal}},\ }\bibfield  {title} {\bibinfo {title} {Theory of the skyrmion,
  meron, antiskyrmion, and antimeron in chiral magnets},\ }\href
  {https://doi.org/10.1103/PhysRevResearch.1.033109} {\bibfield  {journal}
  {\bibinfo  {journal} {Phys. Rev. Res.}\ }\textbf {\bibinfo {volume} {1}},\
  \bibinfo {pages} {033109} (\bibinfo {year} {2019})}\BibitemShut {NoStop}%
\bibitem [{\citenamefont {Meyer}\ \emph {et~al.}(2019)\citenamefont {Meyer},
  \citenamefont {Perini}, \citenamefont {von Malottki}, \citenamefont
  {Kubetzka}, \citenamefont {Wiesendanger}, \citenamefont {von Bergmann},\ and\
  \citenamefont {Heinze}}]{Meyer2019}%
  \BibitemOpen
  \bibfield  {author} {\bibinfo {author} {\bibfnamefont {S.}~\bibnamefont
  {Meyer}}, \bibinfo {author} {\bibfnamefont {M.}~\bibnamefont {Perini}},
  \bibinfo {author} {\bibfnamefont {S.}~\bibnamefont {von Malottki}}, \bibinfo
  {author} {\bibfnamefont {A.}~\bibnamefont {Kubetzka}}, \bibinfo {author}
  {\bibfnamefont {R.}~\bibnamefont {Wiesendanger}}, \bibinfo {author}
  {\bibfnamefont {K.}~\bibnamefont {von Bergmann}},\ and\ \bibinfo {author}
  {\bibfnamefont {S.}~\bibnamefont {Heinze}},\ }\bibfield  {title} {\bibinfo
  {title} {Isolated zero field sub-10 nm skyrmions in ultrathin co films},\
  }\href {https://doi.org/10.1038/s41467-019-11831-4} {\bibfield  {journal}
  {\bibinfo  {journal} {Nature Communications}\ }\textbf {\bibinfo {volume}
  {10}},\ \bibinfo {pages} {3823} (\bibinfo {year} {2019})}\BibitemShut
  {NoStop}%
\bibitem [{\citenamefont {M\"unzer}\ \emph {et~al.}(2010)\citenamefont
  {M\"unzer}, \citenamefont {Neubauer}, \citenamefont {Adams}, \citenamefont
  {M\"uhlbauer}, \citenamefont {Franz}, \citenamefont {Jonietz}, \citenamefont
  {Georgii}, \citenamefont {B\"oni}, \citenamefont {Pedersen}, \citenamefont
  {Schmidt}, \citenamefont {Rosch},\ and\ \citenamefont
  {Pfleiderer}}]{Munzer2010}%
  \BibitemOpen
  \bibfield  {author} {\bibinfo {author} {\bibfnamefont {W.}~\bibnamefont
  {M\"unzer}}, \bibinfo {author} {\bibfnamefont {A.}~\bibnamefont {Neubauer}},
  \bibinfo {author} {\bibfnamefont {T.}~\bibnamefont {Adams}}, \bibinfo
  {author} {\bibfnamefont {S.}~\bibnamefont {M\"uhlbauer}}, \bibinfo {author}
  {\bibfnamefont {C.}~\bibnamefont {Franz}}, \bibinfo {author} {\bibfnamefont
  {F.}~\bibnamefont {Jonietz}}, \bibinfo {author} {\bibfnamefont
  {R.}~\bibnamefont {Georgii}}, \bibinfo {author} {\bibfnamefont
  {P.}~\bibnamefont {B\"oni}}, \bibinfo {author} {\bibfnamefont
  {B.}~\bibnamefont {Pedersen}}, \bibinfo {author} {\bibfnamefont
  {M.}~\bibnamefont {Schmidt}}, \bibinfo {author} {\bibfnamefont
  {A.}~\bibnamefont {Rosch}},\ and\ \bibinfo {author} {\bibfnamefont
  {C.}~\bibnamefont {Pfleiderer}},\ }\bibfield  {title} {\bibinfo {title}
  {Skyrmion lattice in the doped semiconductor
  ${\text{fe}}_{1\ensuremath{-}x}{\text{co}}_{x}\text{Si}$},\ }\href
  {https://doi.org/10.1103/PhysRevB.81.041203} {\bibfield  {journal} {\bibinfo
  {journal} {Phys. Rev. B}\ }\textbf {\bibinfo {volume} {81}},\ \bibinfo
  {pages} {041203} (\bibinfo {year} {2010})}\BibitemShut {NoStop}%
\bibitem [{\citenamefont {Yu}\ \emph {et~al.}(2016)\citenamefont {Yu},
  \citenamefont {Upadhyaya}, \citenamefont {Li}, \citenamefont {Li},
  \citenamefont {Kim}, \citenamefont {Fan}, \citenamefont {Wong}, \citenamefont
  {Tserkovnyak}, \citenamefont {Amiri},\ and\ \citenamefont {Wang}}]{Yu2016}%
  \BibitemOpen
  \bibfield  {author} {\bibinfo {author} {\bibfnamefont {G.}~\bibnamefont
  {Yu}}, \bibinfo {author} {\bibfnamefont {P.}~\bibnamefont {Upadhyaya}},
  \bibinfo {author} {\bibfnamefont {X.}~\bibnamefont {Li}}, \bibinfo {author}
  {\bibfnamefont {W.}~\bibnamefont {Li}}, \bibinfo {author} {\bibfnamefont
  {S.~K.}\ \bibnamefont {Kim}}, \bibinfo {author} {\bibfnamefont
  {Y.}~\bibnamefont {Fan}}, \bibinfo {author} {\bibfnamefont {K.~L.}\
  \bibnamefont {Wong}}, \bibinfo {author} {\bibfnamefont {Y.}~\bibnamefont
  {Tserkovnyak}}, \bibinfo {author} {\bibfnamefont {P.~K.}\ \bibnamefont
  {Amiri}},\ and\ \bibinfo {author} {\bibfnamefont {K.~L.}\ \bibnamefont
  {Wang}},\ }\bibfield  {title} {\bibinfo {title} {Room-temperature creation
  and spin--orbit torque manipulation of skyrmions in thin films with
  engineered asymmetry},\ }\href {https://doi.org/10.1021/acs.nanolett.5b05257}
  {\bibfield  {journal} {\bibinfo  {journal} {Nano Letters}\ }\textbf {\bibinfo
  {volume} {16}},\ \bibinfo {pages} {1981} (\bibinfo {year}
  {2016})}\BibitemShut {NoStop}%
\bibitem [{\citenamefont {Romming}\ \emph {et~al.}(2013)\citenamefont
  {Romming}, \citenamefont {Hanneken}, \citenamefont {Menzel}, \citenamefont
  {Bickel}, \citenamefont {Wolter}, \citenamefont {von Bergmann}, \citenamefont
  {Kubetzka},\ and\ \citenamefont {Wiesendanger}}]{Romming2013}%
  \BibitemOpen
  \bibfield  {author} {\bibinfo {author} {\bibfnamefont {N.}~\bibnamefont
  {Romming}}, \bibinfo {author} {\bibfnamefont {C.}~\bibnamefont {Hanneken}},
  \bibinfo {author} {\bibfnamefont {M.}~\bibnamefont {Menzel}}, \bibinfo
  {author} {\bibfnamefont {J.~E.}\ \bibnamefont {Bickel}}, \bibinfo {author}
  {\bibfnamefont {B.}~\bibnamefont {Wolter}}, \bibinfo {author} {\bibfnamefont
  {K.}~\bibnamefont {von Bergmann}}, \bibinfo {author} {\bibfnamefont
  {A.}~\bibnamefont {Kubetzka}},\ and\ \bibinfo {author} {\bibfnamefont
  {R.}~\bibnamefont {Wiesendanger}},\ }\bibfield  {title} {\bibinfo {title}
  {Writing and deleting single magnetic skyrmions},\ }\href
  {https://doi.org/10.1126/science.1240573} {\bibfield  {journal} {\bibinfo
  {journal} {Science}\ }\textbf {\bibinfo {volume} {341}},\ \bibinfo {pages}
  {636} (\bibinfo {year} {2013})}\BibitemShut {NoStop}%
\bibitem [{\citenamefont {Yu}\ \emph {et~al.}(2010)\citenamefont {Yu},
  \citenamefont {Kanazawa}, \citenamefont {Onose}, \citenamefont {Kimoto},
  \citenamefont {Zhang}, \citenamefont {Ishiwata}, \citenamefont {Matsui},\
  and\ \citenamefont {Tokura}}]{Yu2010}%
  \BibitemOpen
  \bibfield  {author} {\bibinfo {author} {\bibfnamefont {X.~Z.}\ \bibnamefont
  {Yu}}, \bibinfo {author} {\bibfnamefont {N.}~\bibnamefont {Kanazawa}},
  \bibinfo {author} {\bibfnamefont {Y.}~\bibnamefont {Onose}}, \bibinfo
  {author} {\bibfnamefont {K.}~\bibnamefont {Kimoto}}, \bibinfo {author}
  {\bibfnamefont {W.~Z.}\ \bibnamefont {Zhang}}, \bibinfo {author}
  {\bibfnamefont {S.}~\bibnamefont {Ishiwata}}, \bibinfo {author}
  {\bibfnamefont {Y.}~\bibnamefont {Matsui}},\ and\ \bibinfo {author}
  {\bibfnamefont {Y.}~\bibnamefont {Tokura}},\ }\bibfield  {title} {\bibinfo
  {title} {Near room-temperature formation of a skyrmion crystal in thin-films
  of the helimagnet fege},\ }\href {https://doi.org/10.1038/nmat2916}
  {\bibfield  {journal} {\bibinfo  {journal} {Nature Materials}\ }\textbf
  {\bibinfo {volume} {10}},\ \bibinfo {pages} {106} (\bibinfo {year}
  {2010})}\BibitemShut {NoStop}%
\bibitem [{\citenamefont {Herv{\'e}}\ \emph {et~al.}(2018)\citenamefont
  {Herv{\'e}}, \citenamefont {Dup{\'e}}, \citenamefont {Lopes}, \citenamefont
  {B{\"o}ttcher}, \citenamefont {Martins}, \citenamefont {Balashov},
  \citenamefont {Gerhard}, \citenamefont {Sinova},\ and\ \citenamefont
  {Wulfhekel}}]{Herve2018}%
  \BibitemOpen
  \bibfield  {author} {\bibinfo {author} {\bibfnamefont {M.}~\bibnamefont
  {Herv{\'e}}}, \bibinfo {author} {\bibfnamefont {B.}~\bibnamefont {Dup{\'e}}},
  \bibinfo {author} {\bibfnamefont {R.}~\bibnamefont {Lopes}}, \bibinfo
  {author} {\bibfnamefont {M.}~\bibnamefont {B{\"o}ttcher}}, \bibinfo {author}
  {\bibfnamefont {M.~D.}\ \bibnamefont {Martins}}, \bibinfo {author}
  {\bibfnamefont {T.}~\bibnamefont {Balashov}}, \bibinfo {author}
  {\bibfnamefont {L.}~\bibnamefont {Gerhard}}, \bibinfo {author} {\bibfnamefont
  {J.}~\bibnamefont {Sinova}},\ and\ \bibinfo {author} {\bibfnamefont
  {W.}~\bibnamefont {Wulfhekel}},\ }\bibfield  {title} {\bibinfo {title}
  {Stabilizing spin spirals and isolated skyrmions at low magnetic field
  exploiting vanishing magnetic anisotropy},\ }\href
  {https://doi.org/10.1038/s41467-018-03240-w} {\bibfield  {journal} {\bibinfo
  {journal} {Nature Communications}\ }\textbf {\bibinfo {volume} {9}},\
  \bibinfo {pages} {1015} (\bibinfo {year} {2018})}\BibitemShut {NoStop}%
\bibitem [{\citenamefont {Yu}\ \emph {et~al.}(2018{\natexlab{a}})\citenamefont
  {Yu}, \citenamefont {Morikawa}, \citenamefont {Yokouchi}, \citenamefont
  {Shibata}, \citenamefont {Kanazawa}, \citenamefont {Kagawa}, \citenamefont
  {Arima},\ and\ \citenamefont {Tokura}}]{Yu2018}%
  \BibitemOpen
  \bibfield  {author} {\bibinfo {author} {\bibfnamefont {X.}~\bibnamefont
  {Yu}}, \bibinfo {author} {\bibfnamefont {D.}~\bibnamefont {Morikawa}},
  \bibinfo {author} {\bibfnamefont {T.}~\bibnamefont {Yokouchi}}, \bibinfo
  {author} {\bibfnamefont {K.}~\bibnamefont {Shibata}}, \bibinfo {author}
  {\bibfnamefont {N.}~\bibnamefont {Kanazawa}}, \bibinfo {author}
  {\bibfnamefont {F.}~\bibnamefont {Kagawa}}, \bibinfo {author} {\bibfnamefont
  {T.-h.}\ \bibnamefont {Arima}},\ and\ \bibinfo {author} {\bibfnamefont
  {Y.}~\bibnamefont {Tokura}},\ }\bibfield  {title} {\bibinfo {title}
  {Aggregation and collapse dynamics of skyrmions in a non-equilibrium state},\
  }\href {https://doi.org/10.1038/s41567-018-0155-3} {\bibfield  {journal}
  {\bibinfo  {journal} {Nature Physics}\ }\textbf {\bibinfo {volume} {14}},\
  \bibinfo {pages} {832} (\bibinfo {year} {2018}{\natexlab{a}})}\BibitemShut
  {NoStop}%
\bibitem [{\citenamefont {Yu}\ \emph {et~al.}(2018{\natexlab{b}})\citenamefont
  {Yu}, \citenamefont {Koshibae}, \citenamefont {Tokunaga}, \citenamefont
  {Shibata}, \citenamefont {Taguchi}, \citenamefont {Nagaosa},\ and\
  \citenamefont {Tokura}}]{Yu2018564}%
  \BibitemOpen
  \bibfield  {author} {\bibinfo {author} {\bibfnamefont {X.~Z.}\ \bibnamefont
  {Yu}}, \bibinfo {author} {\bibfnamefont {W.}~\bibnamefont {Koshibae}},
  \bibinfo {author} {\bibfnamefont {Y.}~\bibnamefont {Tokunaga}}, \bibinfo
  {author} {\bibfnamefont {K.}~\bibnamefont {Shibata}}, \bibinfo {author}
  {\bibfnamefont {Y.}~\bibnamefont {Taguchi}}, \bibinfo {author} {\bibfnamefont
  {N.}~\bibnamefont {Nagaosa}},\ and\ \bibinfo {author} {\bibfnamefont
  {Y.}~\bibnamefont {Tokura}},\ }\bibfield  {title} {\bibinfo {title}
  {Transformation between meron and skyrmion topological spin textures in a
  chiral magnet},\ }\href {https://doi.org/10.1038/s41586-018-0745-3}
  {\bibfield  {journal} {\bibinfo  {journal} {Nature}\ }\textbf {\bibinfo
  {volume} {564}},\ \bibinfo {pages} {95} (\bibinfo {year}
  {2018}{\natexlab{b}})}\BibitemShut {NoStop}%
\bibitem [{\citenamefont {Gao}\ \emph {et~al.}(2019)\citenamefont {Gao},
  \citenamefont {Je}, \citenamefont {Im}, \citenamefont {Choi}, \citenamefont
  {Yang}, \citenamefont {Li}, \citenamefont {Wang}, \citenamefont {Lee},
  \citenamefont {Han}, \citenamefont {Lee}, \citenamefont {Chao}, \citenamefont
  {Hwang}, \citenamefont {Li},\ and\ \citenamefont {Qiu}}]{Gao2019}%
  \BibitemOpen
  \bibfield  {author} {\bibinfo {author} {\bibfnamefont {N.}~\bibnamefont
  {Gao}}, \bibinfo {author} {\bibfnamefont {S.-G.}\ \bibnamefont {Je}},
  \bibinfo {author} {\bibfnamefont {M.-Y.}\ \bibnamefont {Im}}, \bibinfo
  {author} {\bibfnamefont {J.~W.}\ \bibnamefont {Choi}}, \bibinfo {author}
  {\bibfnamefont {M.}~\bibnamefont {Yang}}, \bibinfo {author} {\bibfnamefont
  {Q.}~\bibnamefont {Li}}, \bibinfo {author} {\bibfnamefont {T.~Y.}\
  \bibnamefont {Wang}}, \bibinfo {author} {\bibfnamefont {S.}~\bibnamefont
  {Lee}}, \bibinfo {author} {\bibfnamefont {H.-S.}\ \bibnamefont {Han}},
  \bibinfo {author} {\bibfnamefont {K.-S.}\ \bibnamefont {Lee}}, \bibinfo
  {author} {\bibfnamefont {W.}~\bibnamefont {Chao}}, \bibinfo {author}
  {\bibfnamefont {C.}~\bibnamefont {Hwang}}, \bibinfo {author} {\bibfnamefont
  {J.}~\bibnamefont {Li}},\ and\ \bibinfo {author} {\bibfnamefont {Z.~Q.}\
  \bibnamefont {Qiu}},\ }\bibfield  {title} {\bibinfo {title} {Creation and
  annihilation of topological meron pairs in in-plane magnetized films},\
  }\href {https://doi.org/10.1038/s41467-019-13642-z} {\bibfield  {journal}
  {\bibinfo  {journal} {Nature Communications}\ }\textbf {\bibinfo {volume}
  {10}},\ \bibinfo {pages} {5603} (\bibinfo {year} {2019})}\BibitemShut
  {NoStop}%
\bibitem [{\citenamefont {Nayak}\ \emph {et~al.}(2017)\citenamefont {Nayak},
  \citenamefont {Kumar}, \citenamefont {Ma}, \citenamefont {Werner},
  \citenamefont {Pippel}, \citenamefont {Sahoo}, \citenamefont {Damay},
  \citenamefont {R{\"o}{\ss}ler}, \citenamefont {Felser},\ and\ \citenamefont
  {Parkin}}]{Nayak2017}%
  \BibitemOpen
  \bibfield  {author} {\bibinfo {author} {\bibfnamefont {A.~K.}\ \bibnamefont
  {Nayak}}, \bibinfo {author} {\bibfnamefont {V.}~\bibnamefont {Kumar}},
  \bibinfo {author} {\bibfnamefont {T.}~\bibnamefont {Ma}}, \bibinfo {author}
  {\bibfnamefont {P.}~\bibnamefont {Werner}}, \bibinfo {author} {\bibfnamefont
  {E.}~\bibnamefont {Pippel}}, \bibinfo {author} {\bibfnamefont
  {R.}~\bibnamefont {Sahoo}}, \bibinfo {author} {\bibfnamefont
  {F.}~\bibnamefont {Damay}}, \bibinfo {author} {\bibfnamefont {U.~K.}\
  \bibnamefont {R{\"o}{\ss}ler}}, \bibinfo {author} {\bibfnamefont
  {C.}~\bibnamefont {Felser}},\ and\ \bibinfo {author} {\bibfnamefont
  {S.~S.~P.}\ \bibnamefont {Parkin}},\ }\bibfield  {title} {\bibinfo {title}
  {Magnetic antiskyrmions above room temperature in tetragonal heusler
  materials},\ }\href {https://doi.org/10.1038/nature23466} {\bibfield
  {journal} {\bibinfo  {journal} {Nature}\ }\textbf {\bibinfo {volume} {548}},\
  \bibinfo {pages} {561} (\bibinfo {year} {2017})}\BibitemShut {NoStop}%
\bibitem [{\citenamefont {Hayami}\ and\ \citenamefont
  {Yambe}(2021)}]{Hayami2021}%
  \BibitemOpen
  \bibfield  {author} {\bibinfo {author} {\bibfnamefont {S.}~\bibnamefont
  {Hayami}}\ and\ \bibinfo {author} {\bibfnamefont {R.}~\bibnamefont {Yambe}},\
  }\bibfield  {title} {\bibinfo {title} {Meron-antimeron crystals in
  noncentrosymmetric itinerant magnets on a triangular lattice},\ }\href
  {https://doi.org/10.1103/PhysRevB.104.094425} {\bibfield  {journal} {\bibinfo
   {journal} {Phys. Rev. B}\ }\textbf {\bibinfo {volume} {104}},\ \bibinfo
  {pages} {094425} (\bibinfo {year} {2021})}\BibitemShut {NoStop}%
\bibitem [{\citenamefont {Nagaosa}\ and\ \citenamefont
  {Tokura}(2013)}]{Nagaosa2013}%
  \BibitemOpen
  \bibfield  {author} {\bibinfo {author} {\bibfnamefont {N.}~\bibnamefont
  {Nagaosa}}\ and\ \bibinfo {author} {\bibfnamefont {Y.}~\bibnamefont
  {Tokura}},\ }\bibfield  {title} {\bibinfo {title} {Topological properties and
  dynamics of magnetic skyrmions},\ }\href
  {https://doi.org/10.1038/nnano.2013.243} {\bibfield  {journal} {\bibinfo
  {journal} {Nature Nanotechnology}\ }\textbf {\bibinfo {volume} {8}},\
  \bibinfo {pages} {899} (\bibinfo {year} {2013})}\BibitemShut {NoStop}%
\bibitem [{\citenamefont {Soumyanarayanan}\ \emph {et~al.}(2017)\citenamefont
  {Soumyanarayanan}, \citenamefont {Raju}, \citenamefont {Gonzalez~Oyarce},
  \citenamefont {Tan}, \citenamefont {Im}, \citenamefont {Petrovic},
  \citenamefont {Ho}, \citenamefont {Khoo}, \citenamefont {Tran}, \citenamefont
  {Gan}, \citenamefont {Ernult},\ and\ \citenamefont
  {Panagopoulos}}]{Soumyanarayanan2017}%
  \BibitemOpen
  \bibfield  {author} {\bibinfo {author} {\bibfnamefont {A.}~\bibnamefont
  {Soumyanarayanan}}, \bibinfo {author} {\bibfnamefont {M.}~\bibnamefont
  {Raju}}, \bibinfo {author} {\bibfnamefont {A.~L.}\ \bibnamefont
  {Gonzalez~Oyarce}}, \bibinfo {author} {\bibfnamefont {A.~K.~C.}\ \bibnamefont
  {Tan}}, \bibinfo {author} {\bibfnamefont {M.-Y.}\ \bibnamefont {Im}},
  \bibinfo {author} {\bibfnamefont {A.~.~P.}\ \bibnamefont {Petrovic}},
  \bibinfo {author} {\bibfnamefont {P.}~\bibnamefont {Ho}}, \bibinfo {author}
  {\bibfnamefont {K.~H.}\ \bibnamefont {Khoo}}, \bibinfo {author}
  {\bibfnamefont {M.}~\bibnamefont {Tran}}, \bibinfo {author} {\bibfnamefont
  {C.~K.}\ \bibnamefont {Gan}}, \bibinfo {author} {\bibfnamefont
  {F.}~\bibnamefont {Ernult}},\ and\ \bibinfo {author} {\bibfnamefont
  {C.}~\bibnamefont {Panagopoulos}},\ }\bibfield  {title} {\bibinfo {title}
  {Tunable room-temperature magnetic skyrmions in ir/fe/co/pt multilayers},\
  }\href {https://doi.org/10.1038/nmat4934} {\bibfield  {journal} {\bibinfo
  {journal} {Nature Materials}\ }\textbf {\bibinfo {volume} {16}},\ \bibinfo
  {pages} {898} (\bibinfo {year} {2017})}\BibitemShut {NoStop}%
\bibitem [{\citenamefont {Bera}\ and\ \citenamefont
  {Mandal}(2021)}]{sandip2020}%
  \BibitemOpen
  \bibfield  {author} {\bibinfo {author} {\bibfnamefont {S.}~\bibnamefont
  {Bera}}\ and\ \bibinfo {author} {\bibfnamefont {S.~S.}\ \bibnamefont
  {Mandal}},\ }\bibfield  {title} {\bibinfo {title} {Skyrmions at vanishingly
  small dzyaloshinskii-moriya interaction or zero magnetic field},\ }\href
  {http://iopscience.iop.org/article/10.1088/1361-648X/abf783} {\bibfield
  {journal} {\bibinfo  {journal} {Journal of Physics: Condensed Matter}\ }
  (\bibinfo {year} {2021})}\BibitemShut {NoStop}%
\bibitem [{\citenamefont {Banerjee}\ \emph {et~al.}(2014)\citenamefont
  {Banerjee}, \citenamefont {Rowland}, \citenamefont {Erten},\ and\
  \citenamefont {Randeria}}]{Banerjee2014}%
  \BibitemOpen
  \bibfield  {author} {\bibinfo {author} {\bibfnamefont {S.}~\bibnamefont
  {Banerjee}}, \bibinfo {author} {\bibfnamefont {J.}~\bibnamefont {Rowland}},
  \bibinfo {author} {\bibfnamefont {O.}~\bibnamefont {Erten}},\ and\ \bibinfo
  {author} {\bibfnamefont {M.}~\bibnamefont {Randeria}},\ }\bibfield  {title}
  {\bibinfo {title} {Enhanced stability of skyrmions in two-dimensional chiral
  magnets with rashba spin-orbit coupling},\ }\href
  {https://doi.org/10.1103/PhysRevX.4.031045} {\bibfield  {journal} {\bibinfo
  {journal} {Phys. Rev. X}\ }\textbf {\bibinfo {volume} {4}},\ \bibinfo {pages}
  {031045} (\bibinfo {year} {2014})}\BibitemShut {NoStop}%
\bibitem [{\citenamefont {Nandy}\ \emph {et~al.}(2016)\citenamefont {Nandy},
  \citenamefont {Kiselev},\ and\ \citenamefont {Bl\"ugel}}]{Nandy2016}%
  \BibitemOpen
  \bibfield  {author} {\bibinfo {author} {\bibfnamefont {A.~K.}\ \bibnamefont
  {Nandy}}, \bibinfo {author} {\bibfnamefont {N.~S.}\ \bibnamefont {Kiselev}},\
  and\ \bibinfo {author} {\bibfnamefont {S.}~\bibnamefont {Bl\"ugel}},\
  }\bibfield  {title} {\bibinfo {title} {Interlayer exchange coupling: A
  general scheme turning chiral magnets into magnetic multilayers carrying
  atomic-scale skyrmions},\ }\href
  {https://doi.org/10.1103/PhysRevLett.116.177202} {\bibfield  {journal}
  {\bibinfo  {journal} {Phys. Rev. Lett.}\ }\textbf {\bibinfo {volume} {116}},\
  \bibinfo {pages} {177202} (\bibinfo {year} {2016})}\BibitemShut {NoStop}%
\bibitem [{\citenamefont {Parkin}\ \emph {et~al.}(2008)\citenamefont {Parkin},
  \citenamefont {Hayashi},\ and\ \citenamefont {Thomas}}]{Parkin2008}%
  \BibitemOpen
  \bibfield  {author} {\bibinfo {author} {\bibfnamefont {S.~S.~P.}\
  \bibnamefont {Parkin}}, \bibinfo {author} {\bibfnamefont {M.}~\bibnamefont
  {Hayashi}},\ and\ \bibinfo {author} {\bibfnamefont {L.}~\bibnamefont
  {Thomas}},\ }\bibfield  {title} {\bibinfo {title} {Magnetic domain-wall
  racetrack memory},\ }\href {https://doi.org/10.1126/science.1145799}
  {\bibfield  {journal} {\bibinfo  {journal} {Science}\ }\textbf {\bibinfo
  {volume} {320}},\ \bibinfo {pages} {190} (\bibinfo {year}
  {2008})}\BibitemShut {NoStop}%
\bibitem [{\citenamefont {Iwasaki}\ \emph {et~al.}(2013)\citenamefont
  {Iwasaki}, \citenamefont {Mochizuki},\ and\ \citenamefont
  {Nagaosa}}]{Iwasaki2013}%
  \BibitemOpen
  \bibfield  {author} {\bibinfo {author} {\bibfnamefont {J.}~\bibnamefont
  {Iwasaki}}, \bibinfo {author} {\bibfnamefont {M.}~\bibnamefont {Mochizuki}},\
  and\ \bibinfo {author} {\bibfnamefont {N.}~\bibnamefont {Nagaosa}},\
  }\bibfield  {title} {\bibinfo {title} {Current-induced skyrmion dynamics in
  constricted geometries},\ }\href {https://doi.org/10.1038/nnano.2013.176}
  {\bibfield  {journal} {\bibinfo  {journal} {Nature Nanotechnology}\ }\textbf
  {\bibinfo {volume} {8}},\ \bibinfo {pages} {742} (\bibinfo {year}
  {2013})}\BibitemShut {NoStop}%
\bibitem [{\citenamefont {Fert}\ \emph {et~al.}(2013)\citenamefont {Fert},
  \citenamefont {Cros},\ and\ \citenamefont {Sampaio}}]{Fert2013}%
  \BibitemOpen
  \bibfield  {author} {\bibinfo {author} {\bibfnamefont {A.}~\bibnamefont
  {Fert}}, \bibinfo {author} {\bibfnamefont {V.}~\bibnamefont {Cros}},\ and\
  \bibinfo {author} {\bibfnamefont {J.}~\bibnamefont {Sampaio}},\ }\bibfield
  {title} {\bibinfo {title} {Skyrmions on the track},\ }\href
  {https://doi.org/10.1038/nnano.2013.29} {\bibfield  {journal} {\bibinfo
  {journal} {Nature Nanotechnology}\ }\textbf {\bibinfo {volume} {8}},\
  \bibinfo {pages} {152} (\bibinfo {year} {2013})}\BibitemShut {NoStop}%
\bibitem [{\citenamefont {Yu}\ \emph {et~al.}(2017)\citenamefont {Yu},
  \citenamefont {Tokunaga}, \citenamefont {Taguchi},\ and\ \citenamefont
  {Tokura}}]{Yu2017}%
  \BibitemOpen
  \bibfield  {author} {\bibinfo {author} {\bibfnamefont {X.}~\bibnamefont
  {Yu}}, \bibinfo {author} {\bibfnamefont {Y.}~\bibnamefont {Tokunaga}},
  \bibinfo {author} {\bibfnamefont {Y.}~\bibnamefont {Taguchi}},\ and\ \bibinfo
  {author} {\bibfnamefont {Y.}~\bibnamefont {Tokura}},\ }\bibfield  {title}
  {\bibinfo {title} {Variation of topology in magnetic bubbles in a colossal
  magnetoresistive manganite},\ }\href
  {https://doi.org/https://doi.org/10.1002/adma.201603958} {\bibfield
  {journal} {\bibinfo  {journal} {Advanced Materials}\ }\textbf {\bibinfo
  {volume} {29}},\ \bibinfo {pages} {1603958} (\bibinfo {year}
  {2017})}\BibitemShut {NoStop}%
\bibitem [{\citenamefont {Sampaio}\ \emph {et~al.}(2013)\citenamefont
  {Sampaio}, \citenamefont {Cros}, \citenamefont {Rohart}, \citenamefont
  {Thiaville},\ and\ \citenamefont {Fert}}]{Sampaio2013}%
  \BibitemOpen
  \bibfield  {author} {\bibinfo {author} {\bibfnamefont {J.}~\bibnamefont
  {Sampaio}}, \bibinfo {author} {\bibfnamefont {V.}~\bibnamefont {Cros}},
  \bibinfo {author} {\bibfnamefont {S.}~\bibnamefont {Rohart}}, \bibinfo
  {author} {\bibfnamefont {A.}~\bibnamefont {Thiaville}},\ and\ \bibinfo
  {author} {\bibfnamefont {A.}~\bibnamefont {Fert}},\ }\bibfield  {title}
  {\bibinfo {title} {Nucleation, stability and current-induced motion of
  isolated magnetic skyrmions in nanostructures},\ }\href
  {https://doi.org/10.1038/nnano.2013.210} {\bibfield  {journal} {\bibinfo
  {journal} {Nature Nanotechnology}\ }\textbf {\bibinfo {volume} {8}},\
  \bibinfo {pages} {839} (\bibinfo {year} {2013})}\BibitemShut {NoStop}%
\bibitem [{\citenamefont {Jonietz}\ \emph {et~al.}(2010)\citenamefont
  {Jonietz}, \citenamefont {M{\"u}hlbauer}, \citenamefont {Pfleiderer},
  \citenamefont {Neubauer}, \citenamefont {M{\"u}nzer}, \citenamefont {Bauer},
  \citenamefont {Adams}, \citenamefont {Georgii}, \citenamefont {B{\"o}ni},
  \citenamefont {Duine}, \citenamefont {Everschor}, \citenamefont {Garst},\
  and\ \citenamefont {Rosch}}]{Jonietz2010}%
  \BibitemOpen
  \bibfield  {author} {\bibinfo {author} {\bibfnamefont {F.}~\bibnamefont
  {Jonietz}}, \bibinfo {author} {\bibfnamefont {S.}~\bibnamefont
  {M{\"u}hlbauer}}, \bibinfo {author} {\bibfnamefont {C.}~\bibnamefont
  {Pfleiderer}}, \bibinfo {author} {\bibfnamefont {A.}~\bibnamefont
  {Neubauer}}, \bibinfo {author} {\bibfnamefont {W.}~\bibnamefont
  {M{\"u}nzer}}, \bibinfo {author} {\bibfnamefont {A.}~\bibnamefont {Bauer}},
  \bibinfo {author} {\bibfnamefont {T.}~\bibnamefont {Adams}}, \bibinfo
  {author} {\bibfnamefont {R.}~\bibnamefont {Georgii}}, \bibinfo {author}
  {\bibfnamefont {P.}~\bibnamefont {B{\"o}ni}}, \bibinfo {author}
  {\bibfnamefont {R.~A.}\ \bibnamefont {Duine}}, \bibinfo {author}
  {\bibfnamefont {K.}~\bibnamefont {Everschor}}, \bibinfo {author}
  {\bibfnamefont {M.}~\bibnamefont {Garst}},\ and\ \bibinfo {author}
  {\bibfnamefont {A.}~\bibnamefont {Rosch}},\ }\bibfield  {title} {\bibinfo
  {title} {Spin transfer torques in mnsi at ultralow current densities},\
  }\href {https://doi.org/10.1126/science.1195709} {\bibfield  {journal}
  {\bibinfo  {journal} {Science}\ }\textbf {\bibinfo {volume} {330}},\ \bibinfo
  {pages} {1648} (\bibinfo {year} {2010})}\BibitemShut {NoStop}%
\bibitem [{\citenamefont {Huang}\ and\ \citenamefont
  {Chien}(2012)}]{Huang2012}%
  \BibitemOpen
  \bibfield  {author} {\bibinfo {author} {\bibfnamefont {S.~X.}\ \bibnamefont
  {Huang}}\ and\ \bibinfo {author} {\bibfnamefont {C.~L.}\ \bibnamefont
  {Chien}},\ }\bibfield  {title} {\bibinfo {title} {Extended skyrmion phase in
  epitaxial $\mathrm{FeGe}(111)$ thin films},\ }\href
  {https://doi.org/10.1103/PhysRevLett.108.267201} {\bibfield  {journal}
  {\bibinfo  {journal} {Phys. Rev. Lett.}\ }\textbf {\bibinfo {volume} {108}},\
  \bibinfo {pages} {267201} (\bibinfo {year} {2012})}\BibitemShut {NoStop}%
\bibitem [{\citenamefont {Li}\ \emph {et~al.}(2013)\citenamefont {Li},
  \citenamefont {Kanazawa}, \citenamefont {Yu}, \citenamefont {Tsukazaki},
  \citenamefont {Kawasaki}, \citenamefont {Ichikawa}, \citenamefont {Jin},
  \citenamefont {Kagawa},\ and\ \citenamefont {Tokura}}]{Li2013}%
  \BibitemOpen
  \bibfield  {author} {\bibinfo {author} {\bibfnamefont {Y.}~\bibnamefont
  {Li}}, \bibinfo {author} {\bibfnamefont {N.}~\bibnamefont {Kanazawa}},
  \bibinfo {author} {\bibfnamefont {X.~Z.}\ \bibnamefont {Yu}}, \bibinfo
  {author} {\bibfnamefont {A.}~\bibnamefont {Tsukazaki}}, \bibinfo {author}
  {\bibfnamefont {M.}~\bibnamefont {Kawasaki}}, \bibinfo {author}
  {\bibfnamefont {M.}~\bibnamefont {Ichikawa}}, \bibinfo {author}
  {\bibfnamefont {X.~F.}\ \bibnamefont {Jin}}, \bibinfo {author} {\bibfnamefont
  {F.}~\bibnamefont {Kagawa}},\ and\ \bibinfo {author} {\bibfnamefont
  {Y.}~\bibnamefont {Tokura}},\ }\bibfield  {title} {\bibinfo {title} {Robust
  formation of skyrmions and topological hall effect anomaly in epitaxial thin
  films of mnsi},\ }\href {https://doi.org/10.1103/PhysRevLett.110.117202}
  {\bibfield  {journal} {\bibinfo  {journal} {Phys. Rev. Lett.}\ }\textbf
  {\bibinfo {volume} {110}},\ \bibinfo {pages} {117202} (\bibinfo {year}
  {2013})}\BibitemShut {NoStop}%
\bibitem [{\citenamefont {Porter}\ \emph {et~al.}(2014)\citenamefont {Porter},
  \citenamefont {Gartside},\ and\ \citenamefont {Marrows}}]{Porter2014}%
  \BibitemOpen
  \bibfield  {author} {\bibinfo {author} {\bibfnamefont {N.~A.}\ \bibnamefont
  {Porter}}, \bibinfo {author} {\bibfnamefont {J.~C.}\ \bibnamefont
  {Gartside}},\ and\ \bibinfo {author} {\bibfnamefont {C.~H.}\ \bibnamefont
  {Marrows}},\ }\bibfield  {title} {\bibinfo {title} {Scattering mechanisms in
  textured fege thin films: Magnetoresistance and the anomalous hall effect},\
  }\href {https://doi.org/10.1103/PhysRevB.90.024403} {\bibfield  {journal}
  {\bibinfo  {journal} {Phys. Rev. B}\ }\textbf {\bibinfo {volume} {90}},\
  \bibinfo {pages} {024403} (\bibinfo {year} {2014})}\BibitemShut {NoStop}%
\bibitem [{\citenamefont {Yokouchi}\ \emph {et~al.}(2014)\citenamefont
  {Yokouchi}, \citenamefont {Kanazawa}, \citenamefont {Tsukazaki},
  \citenamefont {Kozuka}, \citenamefont {Kawasaki}, \citenamefont {Ichikawa},
  \citenamefont {Kagawa},\ and\ \citenamefont {Tokura}}]{Yokouchi2014}%
  \BibitemOpen
  \bibfield  {author} {\bibinfo {author} {\bibfnamefont {T.}~\bibnamefont
  {Yokouchi}}, \bibinfo {author} {\bibfnamefont {N.}~\bibnamefont {Kanazawa}},
  \bibinfo {author} {\bibfnamefont {A.}~\bibnamefont {Tsukazaki}}, \bibinfo
  {author} {\bibfnamefont {Y.}~\bibnamefont {Kozuka}}, \bibinfo {author}
  {\bibfnamefont {M.}~\bibnamefont {Kawasaki}}, \bibinfo {author}
  {\bibfnamefont {M.}~\bibnamefont {Ichikawa}}, \bibinfo {author}
  {\bibfnamefont {F.}~\bibnamefont {Kagawa}},\ and\ \bibinfo {author}
  {\bibfnamefont {Y.}~\bibnamefont {Tokura}},\ }\bibfield  {title} {\bibinfo
  {title} {Stability of two-dimensional skyrmions in thin films of
  mn${}_{1\ensuremath{-}x}$fe${}_{x}$si investigated by the topological hall
  effect},\ }\href {https://doi.org/10.1103/PhysRevB.89.064416} {\bibfield
  {journal} {\bibinfo  {journal} {Phys. Rev. B}\ }\textbf {\bibinfo {volume}
  {89}},\ \bibinfo {pages} {064416} (\bibinfo {year} {2014})}\BibitemShut
  {NoStop}%
\bibitem [{\citenamefont {Göbel}\ \emph {et~al.}(2021)\citenamefont {Göbel},
  \citenamefont {Mertig},\ and\ \citenamefont {Tretiakov}}]{Gobel2021}%
  \BibitemOpen
  \bibfield  {author} {\bibinfo {author} {\bibfnamefont {B.}~\bibnamefont
  {Göbel}}, \bibinfo {author} {\bibfnamefont {I.}~\bibnamefont {Mertig}},\
  and\ \bibinfo {author} {\bibfnamefont {O.~A.}\ \bibnamefont {Tretiakov}},\
  }\bibfield  {title} {\bibinfo {title} {Beyond skyrmions: Review and
  perspectives of alternative magnetic quasiparticles},\ }\href
  {https://doi.org/https://doi.org/10.1016/j.physrep.2020.10.001} {\bibfield
  {journal} {\bibinfo  {journal} {Physics Reports}\ }\textbf {\bibinfo {volume}
  {895}},\ \bibinfo {pages} {1} (\bibinfo {year} {2021})}\BibitemShut {NoStop}%
\bibitem [{\citenamefont {Bera}\ and\ \citenamefont
  {Mandal}(2020)}]{sandip2020l}%
  \BibitemOpen
  \bibfield  {author} {\bibinfo {author} {\bibfnamefont {S.}~\bibnamefont
  {Bera}}\ and\ \bibinfo {author} {\bibfnamefont {S.~S.}\ \bibnamefont
  {Mandal}},\ }\bibfield  {title} {\bibinfo {title} {Length-scale independent
  skyrmion and meron hall angles},\ }\href
  {https://doi.org/10.1088/1361-648X/abd424} {\bibfield  {journal} {\bibinfo
  {journal} {Journal of Physics: Condensed Matter}\ }\textbf {\bibinfo {volume}
  {33}},\ \bibinfo {pages} {115801} (\bibinfo {year} {2020})}\BibitemShut
  {NoStop}%
\bibitem [{\citenamefont {Shigeto}\ \emph {et~al.}(2002)\citenamefont
  {Shigeto}, \citenamefont {Okuno}, \citenamefont {Mibu}, \citenamefont
  {Shinjo},\ and\ \citenamefont {Ono}}]{Shigeto2002}%
  \BibitemOpen
  \bibfield  {author} {\bibinfo {author} {\bibfnamefont {K.}~\bibnamefont
  {Shigeto}}, \bibinfo {author} {\bibfnamefont {T.}~\bibnamefont {Okuno}},
  \bibinfo {author} {\bibfnamefont {K.}~\bibnamefont {Mibu}}, \bibinfo {author}
  {\bibfnamefont {T.}~\bibnamefont {Shinjo}},\ and\ \bibinfo {author}
  {\bibfnamefont {T.}~\bibnamefont {Ono}},\ }\bibfield  {title} {\bibinfo
  {title} {{Magnetic force microscopy observation of antivortex core with
  perpendicular magnetization in patterned thin film of permalloy}},\ }\href
  {https://doi.org/10.1063/1.1483386} {\bibfield  {journal} {\bibinfo
  {journal} {Applied Physics Letters}\ }\textbf {\bibinfo {volume} {80}},\
  \bibinfo {pages} {4190} (\bibinfo {year} {2002})}\BibitemShut {NoStop}%
\bibitem [{\citenamefont {Chmiel}\ \emph {et~al.}(2018)\citenamefont {Chmiel},
  \citenamefont {Waterfield~Price}, \citenamefont {Johnson}, \citenamefont
  {Lamirand}, \citenamefont {Schad}, \citenamefont {van~der Laan},
  \citenamefont {Harris}, \citenamefont {Irwin}, \citenamefont {Rzchowski},
  \citenamefont {Eom},\ and\ \citenamefont {Radaelli}}]{Chmiel2018}%
  \BibitemOpen
  \bibfield  {author} {\bibinfo {author} {\bibfnamefont {F.~P.}\ \bibnamefont
  {Chmiel}}, \bibinfo {author} {\bibfnamefont {N.}~\bibnamefont
  {Waterfield~Price}}, \bibinfo {author} {\bibfnamefont {R.~D.}\ \bibnamefont
  {Johnson}}, \bibinfo {author} {\bibfnamefont {A.~D.}\ \bibnamefont
  {Lamirand}}, \bibinfo {author} {\bibfnamefont {J.}~\bibnamefont {Schad}},
  \bibinfo {author} {\bibfnamefont {G.}~\bibnamefont {van~der Laan}}, \bibinfo
  {author} {\bibfnamefont {D.~T.}\ \bibnamefont {Harris}}, \bibinfo {author}
  {\bibfnamefont {J.}~\bibnamefont {Irwin}}, \bibinfo {author} {\bibfnamefont
  {M.~S.}\ \bibnamefont {Rzchowski}}, \bibinfo {author} {\bibfnamefont {C.-B.}\
  \bibnamefont {Eom}},\ and\ \bibinfo {author} {\bibfnamefont {P.~G.}\
  \bibnamefont {Radaelli}},\ }\bibfield  {title} {\bibinfo {title} {Observation
  of magnetic vortex pairs at room temperature in a planar $\alpha$-fe2o3/co
  heterostructure},\ }\href {https://doi.org/10.1038/s41563-018-0101-x}
  {\bibfield  {journal} {\bibinfo  {journal} {Nature Materials}\ }\textbf
  {\bibinfo {volume} {17}},\ \bibinfo {pages} {581} (\bibinfo {year}
  {2018})}\BibitemShut {NoStop}%
\bibitem [{\citenamefont {Phatak}\ \emph {et~al.}(2012)\citenamefont {Phatak},
  \citenamefont {Petford-Long},\ and\ \citenamefont {Heinonen}}]{Phatak2012}%
  \BibitemOpen
  \bibfield  {author} {\bibinfo {author} {\bibfnamefont {C.}~\bibnamefont
  {Phatak}}, \bibinfo {author} {\bibfnamefont {A.~K.}\ \bibnamefont
  {Petford-Long}},\ and\ \bibinfo {author} {\bibfnamefont {O.}~\bibnamefont
  {Heinonen}},\ }\bibfield  {title} {\bibinfo {title} {Direct observation of
  unconventional topological spin structure in coupled magnetic discs},\ }\href
  {https://doi.org/10.1103/PhysRevLett.108.067205} {\bibfield  {journal}
  {\bibinfo  {journal} {Phys. Rev. Lett.}\ }\textbf {\bibinfo {volume} {108}},\
  \bibinfo {pages} {067205} (\bibinfo {year} {2012})}\BibitemShut {NoStop}%
\bibitem [{\citenamefont {Heide}\ \emph {et~al.}(2008)\citenamefont {Heide},
  \citenamefont {Bihlmayer},\ and\ \citenamefont {Bl\"ugel}}]{Heide2008}%
  \BibitemOpen
  \bibfield  {author} {\bibinfo {author} {\bibfnamefont {M.}~\bibnamefont
  {Heide}}, \bibinfo {author} {\bibfnamefont {G.}~\bibnamefont {Bihlmayer}},\
  and\ \bibinfo {author} {\bibfnamefont {S.}~\bibnamefont {Bl\"ugel}},\
  }\bibfield  {title} {\bibinfo {title} {Dzyaloshinskii-moriya interaction
  accounting for the orientation of magnetic domains in ultrathin films:
  Fe/w(110)},\ }\href {https://doi.org/10.1103/PhysRevB.78.140403} {\bibfield
  {journal} {\bibinfo  {journal} {Phys. Rev. B}\ }\textbf {\bibinfo {volume}
  {78}},\ \bibinfo {pages} {140403} (\bibinfo {year} {2008})}\BibitemShut
  {NoStop}%
\bibitem [{\citenamefont {Camosi}\ \emph {et~al.}(2017)\citenamefont {Camosi},
  \citenamefont {Rohart}, \citenamefont {Fruchart}, \citenamefont {Pizzini},
  \citenamefont {Belmeguenai}, \citenamefont {Roussign\'e}, \citenamefont
  {Stashkevich}, \citenamefont {Cherif}, \citenamefont {Ranno}, \citenamefont
  {de~Santis},\ and\ \citenamefont {Vogel}}]{Camosi2017}%
  \BibitemOpen
  \bibfield  {author} {\bibinfo {author} {\bibfnamefont {L.}~\bibnamefont
  {Camosi}}, \bibinfo {author} {\bibfnamefont {S.}~\bibnamefont {Rohart}},
  \bibinfo {author} {\bibfnamefont {O.}~\bibnamefont {Fruchart}}, \bibinfo
  {author} {\bibfnamefont {S.}~\bibnamefont {Pizzini}}, \bibinfo {author}
  {\bibfnamefont {M.}~\bibnamefont {Belmeguenai}}, \bibinfo {author}
  {\bibfnamefont {Y.}~\bibnamefont {Roussign\'e}}, \bibinfo {author}
  {\bibfnamefont {A.}~\bibnamefont {Stashkevich}}, \bibinfo {author}
  {\bibfnamefont {S.~M.}\ \bibnamefont {Cherif}}, \bibinfo {author}
  {\bibfnamefont {L.}~\bibnamefont {Ranno}}, \bibinfo {author} {\bibfnamefont
  {M.}~\bibnamefont {de~Santis}},\ and\ \bibinfo {author} {\bibfnamefont
  {J.}~\bibnamefont {Vogel}},\ }\bibfield  {title} {\bibinfo {title}
  {Anisotropic dzyaloshinskii-moriya interaction in ultrathin epitaxial
  au/co/w(110)},\ }\href {https://doi.org/10.1103/PhysRevB.95.214422}
  {\bibfield  {journal} {\bibinfo  {journal} {Phys. Rev. B}\ }\textbf {\bibinfo
  {volume} {95}},\ \bibinfo {pages} {214422} (\bibinfo {year}
  {2017})}\BibitemShut {NoStop}%
\bibitem [{\citenamefont {Liu}\ \emph {et~al.}(2021)\citenamefont {Liu},
  \citenamefont {Zhang}, \citenamefont {Chai},\ and\ \citenamefont
  {Wu}}]{Liu2021}%
  \BibitemOpen
  \bibfield  {author} {\bibinfo {author} {\bibfnamefont {C.~Q.}\ \bibnamefont
  {Liu}}, \bibinfo {author} {\bibfnamefont {Y.~B.}\ \bibnamefont {Zhang}},
  \bibinfo {author} {\bibfnamefont {G.~Z.}\ \bibnamefont {Chai}},\ and\
  \bibinfo {author} {\bibfnamefont {Y.~Z.}\ \bibnamefont {Wu}},\ }\bibfield
  {title} {\bibinfo {title} {{Large anisotropic Dzyaloshinskii–Moriya
  interaction in CoFeB(211)/Pt(110) films}},\ }\bibfield  {journal} {\bibinfo
  {journal} {Applied Physics Letters}\ }\textbf {\bibinfo {volume} {118}},\
  \href {https://doi.org/10.1063/5.0054943} {10.1063/5.0054943} (\bibinfo
  {year} {2021}),\ \bibinfo {note} {262410}\BibitemShut {NoStop}%
\bibitem [{\citenamefont {Shibata}\ \emph {et~al.}(2015)\citenamefont
  {Shibata}, \citenamefont {Iwasaki}, \citenamefont {Kanazawa}, \citenamefont
  {Aizawa}, \citenamefont {Tanigaki}, \citenamefont {Shirai}, \citenamefont
  {Nakajima}, \citenamefont {Kubota}, \citenamefont {Kawasaki}, \citenamefont
  {Park}, \citenamefont {Shindo}, \citenamefont {Nagaosa},\ and\ \citenamefont
  {Tokura}}]{Shibata2015}%
  \BibitemOpen
  \bibfield  {author} {\bibinfo {author} {\bibfnamefont {K.}~\bibnamefont
  {Shibata}}, \bibinfo {author} {\bibfnamefont {J.}~\bibnamefont {Iwasaki}},
  \bibinfo {author} {\bibfnamefont {N.}~\bibnamefont {Kanazawa}}, \bibinfo
  {author} {\bibfnamefont {S.}~\bibnamefont {Aizawa}}, \bibinfo {author}
  {\bibfnamefont {T.}~\bibnamefont {Tanigaki}}, \bibinfo {author}
  {\bibfnamefont {M.}~\bibnamefont {Shirai}}, \bibinfo {author} {\bibfnamefont
  {T.}~\bibnamefont {Nakajima}}, \bibinfo {author} {\bibfnamefont
  {M.}~\bibnamefont {Kubota}}, \bibinfo {author} {\bibfnamefont
  {M.}~\bibnamefont {Kawasaki}}, \bibinfo {author} {\bibfnamefont {H.~S.}\
  \bibnamefont {Park}}, \bibinfo {author} {\bibfnamefont {D.}~\bibnamefont
  {Shindo}}, \bibinfo {author} {\bibfnamefont {N.}~\bibnamefont {Nagaosa}},\
  and\ \bibinfo {author} {\bibfnamefont {Y.}~\bibnamefont {Tokura}},\
  }\bibfield  {title} {\bibinfo {title} {Large anisotropic deformation of
  skyrmions in strained crystal},\ }\href
  {https://doi.org/10.1038/nnano.2015.113} {\bibfield  {journal} {\bibinfo
  {journal} {Nature Nanotechnology}\ }\textbf {\bibinfo {volume} {10}},\
  \bibinfo {pages} {589} (\bibinfo {year} {2015})}\BibitemShut {NoStop}%
\bibitem [{\citenamefont {Hoffmann}\ \emph {et~al.}(2017)\citenamefont
  {Hoffmann}, \citenamefont {Zimmermann}, \citenamefont {M{\"u}ller},
  \citenamefont {Sch{\"u}rhoff}, \citenamefont {Kiselev}, \citenamefont
  {Melcher},\ and\ \citenamefont {Bl{\"u}gel}}]{Hoffmann2017}%
  \BibitemOpen
  \bibfield  {author} {\bibinfo {author} {\bibfnamefont {M.}~\bibnamefont
  {Hoffmann}}, \bibinfo {author} {\bibfnamefont {B.}~\bibnamefont
  {Zimmermann}}, \bibinfo {author} {\bibfnamefont {G.~P.}\ \bibnamefont
  {M{\"u}ller}}, \bibinfo {author} {\bibfnamefont {D.}~\bibnamefont
  {Sch{\"u}rhoff}}, \bibinfo {author} {\bibfnamefont {N.~S.}\ \bibnamefont
  {Kiselev}}, \bibinfo {author} {\bibfnamefont {C.}~\bibnamefont {Melcher}},\
  and\ \bibinfo {author} {\bibfnamefont {S.}~\bibnamefont {Bl{\"u}gel}},\
  }\bibfield  {title} {\bibinfo {title} {Antiskyrmions stabilized at interfaces
  by anisotropic dzyaloshinskii-moriya interactions},\ }\href
  {https://doi.org/10.1038/s41467-017-00313-0} {\bibfield  {journal} {\bibinfo
  {journal} {Nature Communications}\ }\textbf {\bibinfo {volume} {8}},\
  \bibinfo {pages} {308} (\bibinfo {year} {2017})}\BibitemShut {NoStop}%
\bibitem [{\citenamefont {Metropolis}\ \emph {et~al.}(1953)\citenamefont
  {Metropolis}, \citenamefont {Rosenbluth}, \citenamefont {Rosenbluth},
  \citenamefont {Teller},\ and\ \citenamefont {Teller}}]{Metropolis1953}%
  \BibitemOpen
  \bibfield  {author} {\bibinfo {author} {\bibfnamefont {N.}~\bibnamefont
  {Metropolis}}, \bibinfo {author} {\bibfnamefont {A.~W.}\ \bibnamefont
  {Rosenbluth}}, \bibinfo {author} {\bibfnamefont {M.~N.}\ \bibnamefont
  {Rosenbluth}}, \bibinfo {author} {\bibfnamefont {A.~H.}\ \bibnamefont
  {Teller}},\ and\ \bibinfo {author} {\bibfnamefont {E.}~\bibnamefont
  {Teller}},\ }\bibfield  {title} {\bibinfo {title} {Equation of state
  calculations by fast computing machines},\ }\href
  {https://doi.org/10.1063/1.1699114} {\bibfield  {journal} {\bibinfo
  {journal} {The Journal of Chemical Physics}\ }\textbf {\bibinfo {volume}
  {21}},\ \bibinfo {pages} {1087} (\bibinfo {year} {1953})}\BibitemShut
  {NoStop}%
\bibitem [{\citenamefont {Hayami}\ and\ \citenamefont
  {Motome}(2018)}]{Hayami2018}%
  \BibitemOpen
  \bibfield  {author} {\bibinfo {author} {\bibfnamefont {S.}~\bibnamefont
  {Hayami}}\ and\ \bibinfo {author} {\bibfnamefont {Y.}~\bibnamefont
  {Motome}},\ }\bibfield  {title} {\bibinfo {title} {N\'eel- and bloch-type
  magnetic vortices in rashba metals},\ }\href
  {https://doi.org/10.1103/PhysRevLett.121.137202} {\bibfield  {journal}
  {\bibinfo  {journal} {Phys. Rev. Lett.}\ }\textbf {\bibinfo {volume} {121}},\
  \bibinfo {pages} {137202} (\bibinfo {year} {2018})}\BibitemShut {NoStop}%
\bibitem [{\citenamefont {Kanazawa}\ \emph {et~al.}(2011)\citenamefont
  {Kanazawa}, \citenamefont {Onose}, \citenamefont {Arima}, \citenamefont
  {Okuyama}, \citenamefont {Ohoyama}, \citenamefont {Wakimoto}, \citenamefont
  {Kakurai}, \citenamefont {Ishiwata},\ and\ \citenamefont
  {Tokura}}]{Kanazawa2011}%
  \BibitemOpen
  \bibfield  {author} {\bibinfo {author} {\bibfnamefont {N.}~\bibnamefont
  {Kanazawa}}, \bibinfo {author} {\bibfnamefont {Y.}~\bibnamefont {Onose}},
  \bibinfo {author} {\bibfnamefont {T.}~\bibnamefont {Arima}}, \bibinfo
  {author} {\bibfnamefont {D.}~\bibnamefont {Okuyama}}, \bibinfo {author}
  {\bibfnamefont {K.}~\bibnamefont {Ohoyama}}, \bibinfo {author} {\bibfnamefont
  {S.}~\bibnamefont {Wakimoto}}, \bibinfo {author} {\bibfnamefont
  {K.}~\bibnamefont {Kakurai}}, \bibinfo {author} {\bibfnamefont
  {S.}~\bibnamefont {Ishiwata}},\ and\ \bibinfo {author} {\bibfnamefont
  {Y.}~\bibnamefont {Tokura}},\ }\bibfield  {title} {\bibinfo {title} {Large
  topological {H}all effect in a short-period helimagnet mnge},\ }\href
  {https://doi.org/10.1103/PhysRevLett.106.156603} {\bibfield  {journal}
  {\bibinfo  {journal} {Phys. Rev. Lett.}\ }\textbf {\bibinfo {volume} {106}},\
  \bibinfo {pages} {156603} (\bibinfo {year} {2011})}\BibitemShut {NoStop}%
\bibitem [{\citenamefont {Romming}\ \emph {et~al.}(2015)\citenamefont
  {Romming}, \citenamefont {Kubetzka}, \citenamefont {Hanneken}, \citenamefont
  {von Bergmann},\ and\ \citenamefont {Wiesendanger}}]{Romming2015}%
  \BibitemOpen
  \bibfield  {author} {\bibinfo {author} {\bibfnamefont {N.}~\bibnamefont
  {Romming}}, \bibinfo {author} {\bibfnamefont {A.}~\bibnamefont {Kubetzka}},
  \bibinfo {author} {\bibfnamefont {C.}~\bibnamefont {Hanneken}}, \bibinfo
  {author} {\bibfnamefont {K.}~\bibnamefont {von Bergmann}},\ and\ \bibinfo
  {author} {\bibfnamefont {R.}~\bibnamefont {Wiesendanger}},\ }\bibfield
  {title} {\bibinfo {title} {Field-dependent size and shape of single magnetic
  skyrmions},\ }\href {https://doi.org/10.1103/PhysRevLett.114.177203}
  {\bibfield  {journal} {\bibinfo  {journal} {Phys. Rev. Lett.}\ }\textbf
  {\bibinfo {volume} {114}},\ \bibinfo {pages} {177203} (\bibinfo {year}
  {2015})}\BibitemShut {NoStop}%
\end{thebibliography}%
\end{document}